\documentclass[cernpreprint, atlasdraft=false, UKenglish, texmf, orcidlogo]{atlasdoc}
 
\usepackage{atlaspackage}
\usepackage{atlasbiblatex}
\usepackage{multirow}
\usepackage{array}
\usepackage{xcolor}
\newcolumntype{~}{>{\global\let\currentrowstyle\relax}}
\newcolumntype{^}{>{\currentrowstyle}}

\usepackage{atlasphysics}

\graphicspath{{logos/}{figures/}}
 
\usepackage{ANA-EXOT-2020-08-PAPER-defs}

\AtlasTitle{Search for leptoquark pair production decaying into $te^- \bar{t}e^+$ or $t\mu^- \bar{t}\mu^+$ in multi-lepton final states in $pp$ collisions at $\sqrt{s} = 13\,\mathrm{TeV}$ with the ATLAS detector}

\AtlasAbstract{
A search for leptoquark pair production decaying into $te^- \bar{t}e^+$ or $t\mu^- \bar{t}\mu^+$ in final states with multiple leptons is presented. The search is based on a dataset of $pp$ collisions at $\sqrt{s}=13~\tev$ recorded with the ATLAS detector during Run 2 of the Large Hadron Collider, corresponding to an integrated luminosity of 139~fb$^{-1}$. Four signal regions, with the requirement of at least three light leptons (electron or muon) and at least two jets out of which at least one jet is identified as coming from a $b$-hadron, are considered based on the number of leptons of a given flavour. \ The main background processes are estimated using dedicated control regions in a simultaneous fit with the signal regions to data. No excess above the Standard Model background prediction is observed and 95\% confidence level limits on the production cross section times branching ratio are derived as a function of the leptoquark mass. Under the assumption of exclusive decays into $te^{-}$ ($t\mu^{-}$), the corresponding lower limit on the scalar mixed-generation leptoquark mass $m_{\ensuremath{\mathrm{LQ}_{\mathrm{mix}}^{\mathrm{d}}}}$ is at 1.58 (1.59) TeV and on the vector leptoquark mass $m_{\ensuremath{\tilde{U}_1}}$ at 1.67 (1.67) TeV in the minimal coupling scenario and at 1.95 (1.95) TeV in the Yang-Mills scenario.
}
 
\author{The ATLAS Collaboration}
 
\AtlasRefCode{EXOT-2020-08}
 
\PreprintIdNumber{CERN-EP-2023-087}

\AtlasJournal{EPJC}
\AtlasJournalRef{Eur. Phys. J. C 84 (2024) 818}
\AtlasDOI{10.1140/epjc/s10052-024-12975-4}

\hypersetup{pdftitle={ATLAS document},pdfauthor={The ATLAS Collaboration}}
 
\begin{document}
 
\maketitle
 
\tableofcontents

\section{Introduction}

The similarities in the structure of the lepton and quark sectors of the Standard Model (SM) raise the possibility of an existing underlying symmetry connecting the two sectors. Consequently, many extensions of the SM of particle physics contain leptoquarks (LQ)~\cite{pati_salam_colour,georgi_glashow_unification,dimop_suss,techni2, techni3, string, comp}, hypothetical particles that carry non-zero baryon and lepton quantum numbers and are charged under all SM gauge groups.
In particular, LQs are triplets with respect to the strong interaction, and have fractional electric charge. A LQ state can have either spin 0 (scalar LQ) or spin 1 (vector LQ), and both are considered in this paper.
Due to their quantum numbers, LQs couple simultaneously to both quarks and leptons, enabling direct transitions between the two.
Consequently, LQs can mediate processes that violate lepton-flavour universality, and were proposed~\cite{Freytsis:2015qca,Bauer:2015knc,DiLuzio:2017chi,Buttazzo:2017ixm,Cline:2017aed,Gino:2023lhd} as an explanation for measurements of $B$-meson decays that exhibit deviations from the SM predictions~\cite{BaBar:2013mob,Belle:2015qfa,LHCb:2015gmp,Belle:2016dyj,LHCb:2017vlu}.
Contributions from a scalar LQ could also account~\cite{Bauer:2015knc,Crivellin:2019dwb,Gherardi:2020qhc,Lee:2021jdr} for the longstanding discrepancy between the measured and the predicted anomalous magnetic moment of the muon $(g-2)_\mu$~\cite{Muong-2:2006rrc,Muong-2:2021ojo}.
 
In $pp$ collisions at the Large Hadron Collider (LHC), pairs of LQs are mainly produced via gluon--gluon fusion and quark--antiquark annihilation, mediated by the strong interaction. There are also lepton-mediated $t$- and $u$-channel production processes that depend on the coupling strength ($\lambda$) of the quark--lepton--LQ interaction. However, these contributions can usually be neglected in the scenario where LQs couple preferentially to third-generation quarks, since the latter would be required in the initial state.
The $\mathrm{LQ}$ pair-production cross section can therefore, to a very good approximation, be taken to depend only on the assumed value of the LQ mass ($m_\mathrm{LQ}$) for a given LQ spin and centre-of-mass energy~\cite{JBluemlein1997,Borschensky:2020hot}.
Single LQ production in association with a lepton is also possible, but the cross section depends on the strength of the quark--lepton--LQ interaction and it is not considered in this paper.
 
Most searches~\cite{EXOT-2015-19,EXOT-2017-30,SUSY-2018-12,EXOT-2019-13,EXOT-2019-15,SUSY-2018-34,SUSY-2019-18,CMS-B2G-16-028,CMS-EXO-16-016,CMS-EXO-16-023,CMS-EXO-17-003,CMS-EXO-17-009,CMS-EXO-17-016,CMS-EXO-17-029,CMS-EXO-19-015} by the ATLAS and CMS experiments at the LHC have targeted the scenario where LQs only interact with leptons and quarks of the same generation as suggested by the models in Ref.~\cite{Buchmuller:1986zs}. However, the possibility of flavour-off-diagonal couplings should also be explored as part of a broad and exhaustive LQ search programme. While the simultaneous coupling of a LQ to more than one generation of quarks or to more than one generation of leptons would lead to quark and lepton flavour violation ($K - \bar{K}$ mixing, $\mu \to e\gamma$), the coupling of LQ to a quark from one generation and to a lepton from another one can still preserve the flavour symmetries~\cite{Diaz_2017}.
The cross-generational LQ search has gained additional interest given the fact that explanations of the $B$-meson and $(g-2)_\mu$ anomalies require LQs with couplings to quarks of the second or third generation and leptons of the first, second or third generation~\cite{GRELJO2021136554,CFI:2019vlq}.
 
The ATLAS Collaboration has performed searches for pair production of both scalar and vector LQs. The different ATLAS search results are presented as a function of the LQ mass and the branching ratio into charged leptons ($\BR$). In the case of scalar LQs interacting with mixed quark and lepton generation ($\LQmix$), two different types were considered: up-type LQs coupling to a third generation quark (top-quark $t$ or b-quark $b$) and a first or second generation lepton ($\LQmixu \to b\ell^+~ \textrm{or}~ t\bar{\nu}$), with $\ell=e, \mu$, and down-type LQs ($\LQmixd \to t\ell^{-}~ \textrm{or}~ b\nu$).  The $\LQmixu$ and $\LQmixd$ have fractional electric charges of $+2/3e$ and $-1/3e$, respectively. In the case of vector LQ, the iso-singlet LQ ($U_1$) carrying electric charge $+2/3e$ with two different coupling scenarios to the SM gauge bosons was considered: the minimal ($U_1^{\mathrm{min}}$) and the Yang-Mills ($U_1^{\mathrm{YM}}$) coupling scenarios. Using the full Run 2 dataset, corresponding to 139~fb$^{-1}$ of $pp$ collisions at $\sqrt{s}=13~\tev$, these searches have excluded at the 95\% confidence level (CL) masses below $1.8~\tev$ ($1.7~\tev$) for $\LQmixu$~\cite{EXOT-2019-13}, and below $1.48~\tev$ ($1.47~\tev$) for $\LQmixd$~\cite{EXOT-2019-19}, assuming $\BR=1$ into electrons (muons). A dedicated search optimised for $\BR=0.5$~\cite{EXOT-2019-12} has probed a wide range of scenarios, including both $\LQmixu$ and $\LQmixd$, as well as $U_1^{\mathrm{min}}$ and $U_1^{\mathrm{YM}}$. The corresponding mass exclusions for scalar LQs range from $1.37~\tev$ to $1.46~\tev$, whereas for $U_1^{\mathrm{min}}$ and $U_1^{\mathrm{YM}}$ they range from $1.62~\tev$ to $1.98~\tev$. The CMS Collaboration has also published searches for pair production of scalar LQs decaying into a top-quark and an electron or a muon~\cite{CMS:2018oaj,CMS:2022nty}, achieving comparable or lower mass exclusions than ATLAS.
 
\begin{figure}[t]
\centering
\subfloat[]{\includegraphics[width=0.49\textwidth]{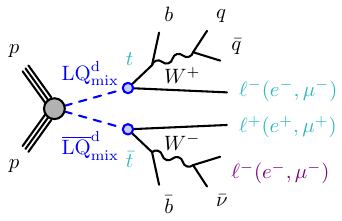}\label{fig:lqlq-tltl-3l}}
\subfloat[]{\includegraphics[width=0.49\textwidth]{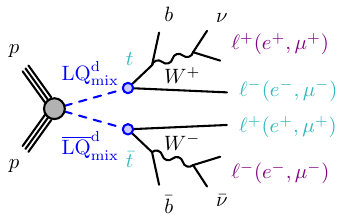}\label{fig:lqlq-tltl-4l}}
\caption{
Signal diagrams of the decay modes of pair-produced scalar leptoquarks $\LQmixd$ targeted in this analysis: (a) \lll and (b) \llll. Decays of the $W$-boson to an electron or a muon of via a $\tau$-lepton ($W \to \tau \nu \to \ell \nu \nu$) are also considered. The vector leptoquarks $\vLQ$ production and decay modes can be described with analogous diagrams. \label{fig:diagrams_with_decay}}
\end{figure}

This paper presents a dedicated search for the pair production of $\LQmixd$ in the $t\ell^- \bar{t}\ell^+$ decay mode (denoted as \tltl\ hereafter), where same-flavour leptons from the $\LQmixd$ decay are considered (denoted as \tete\ and \tmutmu), and the top-quark pair is required to decay into a final state with one or two leptons as illustrated in Figure~\ref{fig:diagrams_with_decay}. This selection is orthogonal to that used by Ref.~\cite{EXOT-2019-19}, where the top-quark pair is required to decay into a final state with jets, and therefore provides additional sensitivity to probe $\LQmixd$. The multi-lepton final state in this analysis takes advantage of the similar top-quark pair branching fraction to final states with at least one lepton compared to those with jets, and of the suppression of SM background processes when selecting events with multiple leptons. Figures~\ref{fig:lqlq-tltl-3l} and~\ref{fig:lqlq-tltl-4l} illustrate the signal processes targeted in this analysis. These decay modes are hereafter indicated without labelling the particle and anti-particle charges. This search uses the full Run~2 dataset of $pp$ collisions at $\sqrt{s}=13~\tev$ recorded with the ATLAS detector and corresponding to an integrated luminosity of 139~fb$^{-1}$.  Events are selected if they have at least two light leptons (electron or muon), and at least two jets, at least one of which must be identified as originating from $B$-hadrons. Three final states defined by the multiplicity of lepton candidates, 2$\ell$ with same electric charge (\llSS), \lll, and \llll, are considered in the analysis. Each of them is split into multiple event categories. Signal-enriched event categories require at least three leptons and exploit the presence of energetic final-state objects, whereas signal-depleted regions are also defined containing \llSS events.
 
The composition of background processes varies among the signal-enriched event categories, however the main background contributions arise from SM processes yielding multiple leptons in the final state, such as $\ttbar$ production in association with a vector boson and diboson production. In those event categories the final discriminating variable used is the scalar sum of the transverse momenta of all selected leptons, the selected jets and the missing transverse momentum. Higher values of this variable are expected for the signal compared to the background. The rest of the event categories are designed to be enriched in the most relevant backgrounds. A maximum-likelihood fit is performed across event categories to search for the signal and constrain several leading backgrounds simultaneously. This search is performed in the $\LQmixd$ mass range between $1~\tev$ and $1.9~\tev$.  The results are also interpreted for the first time in the context of an iso-singlet vector LQ with electric charge $+5/3e$~\cite{Buchmuller:1986zs}. This vector LQ has right-handed coupling to a top quark and a charged lepton and is labelled as $\vLQ$.  The mass range considered is between $1~\tev$ and $2~\tev$ for two different coupling assumptions with other gauge bosons.
 
\section{ATLAS detector}

\newcommand{\AtlasCoordFootnote}{
ATLAS uses a right-handed coordinate system with its origin at the nominal interaction point (IP)
in the centre of the detector and the \(z\)-axis along the beam pipe.
The \(x\)-axis points from the IP to the centre of the LHC ring,
and the \(y\)-axis points upwards.
Cylindrical coordinates \((r,\phi)\) are used in the transverse plane,
\(\phi\) being the azimuthal angle around the \(z\)-axis.
The pseudorapidity is defined in terms of the polar angle \(\theta\) as \(\eta = -\ln \tan(\theta/2)\).
Angular distance is measured in units of \(\Delta R \equiv \sqrt{(\Delta\eta)^{2} + (\Delta\phi)^{2}}\).}
 
The ATLAS detector~\cite{PERF-2007-01} at the LHC covers nearly the entire solid angle around the collision point.\footnote{\AtlasCoordFootnote}
It consists of an inner tracking detector surrounded by a thin superconducting solenoid, electromagnetic and hadron calorimeters,
and a muon spectrometer incorporating three large superconducting air-core toroidal magnets.
 
The inner-detector system (ID) is immersed in a \SI{2}{\tesla} axial magnetic field
and provides charged-particle tracking in the range \(|\eta| < 2.5\).
The high-granularity silicon pixel detector covers the vertex region and typically provides four measurements per track,
the first hit normally being in the insertable B-layer (IBL) installed before Run~2~\cite{ATLAS-TDR-19,PIX-2018-001}.
It is followed by the silicon microstrip tracker (SCT), which usually provides eight measurements per track.
These silicon detectors are complemented by the transition radiation tracker (TRT),
which enables radially extended track reconstruction up to \(|\eta| = 2.0\).
The TRT also provides electron identification information
based on the fraction of hits (typically 30 in total) above a higher energy-deposit threshold corresponding to transition radiation.
 
The calorimeter system covers the pseudorapidity range \(|\eta| < 4.9\).
Within the region \(|\eta|< 3.2\), electromagnetic calorimetry is provided by barrel and
endcap high-granularity lead/liquid-argon (LAr) calorimeters,
with an additional thin LAr presampler covering \(|\eta| < 1.8\)
to correct for energy loss in material upstream of the calorimeters.
Hadron calorimetry is provided by the steel/scintillator-tile calorimeter,
segmented into three barrel structures within \(|\eta| < 1.7\), and two copper/LAr hadron endcap calorimeters.
The solid angle coverage is completed with forward copper/LAr and tungsten/LAr calorimeter modules
optimised for electromagnetic and hadronic energy measurements respectively.
 
The muon spectrometer (MS) comprises separate trigger and
high-precision tracking chambers measuring the deflection of muons in a magnetic field generated by the superconducting air-core toroidal magnets.
The field integral of the toroids ranges between \num{2.0} and \SI{6.0}{\tesla\metre}
across most of the detector.
Three layers of precision chambers, each consisting of layers of monitored drift tubes, cover the region \(|\eta| < 2.7\),
complemented by cathode-strip chambers in the forward region, where the background is highest.
The muon trigger system covers the range \(|\eta| < 2.4\) with resistive-plate chambers in the barrel, and thin-gap chambers in the endcap regions.
 
Interesting events are selected by the first-level trigger system implemented in custom hardware,
followed by selections made by algorithms implemented in software in the high-level trigger~\cite{TRIG-2016-01}.
The first-level trigger accepts events from the \SI{40}{\MHz} bunch crossings at a rate below \SI{100}{\kHz},
which the high-level trigger further reduces in order to record events to disk at about \SI{1}{\kHz}.
 
An extensive software suite~\cite{ATL-SOFT-PUB-2021-001} is used in the reconstruction and analysis of real
and simulated data, in detector operations, and in the trigger and data acquisition systems of the experiment.


\section{Data and simulated event samples}
\label{sec:datamc}

 
A dataset of $pp$ collisions at $\sqrt{s}=13~\tev$ collected by the ATLAS experiment during 2015--2018 and corresponding to an integrated luminosity of $139~\ifb$ is used.
The number of additional $pp$ interactions per bunch crossing (pile-up) in this dataset ranges from about 8 to 70, with an average of 34. Only events recorded under stable beam conditions and for which all relevant detector subsystems were known to be in a good operating condition are used. Events are selected with a combination of lepton triggers, as discussed in Section~\ref{sec:events}.
 
Monte Carlo (MC) simulation samples were produced for the different signal and background processes.
All simulated samples, except those produced with the \textsc{Sherpa}~\cite{Gleisberg:2008ta} event generator, utilised \textsc{EvtGen} 1.2.0~\cite{Lange:2001uf} to model the decays of heavy-flavour hadrons.
Pile-up was modelled by overlaying the simulated hard-scattering event with inelastic $pp$ events generated with \PYTHIA[8.186]~\cite{Sjostrand:2007gs} using the \NNPDF[2.3lo] set of parton distribution functions (PDF)~\cite{Ball:2012cx} and the A3 set of tuned parameters~\cite{ATL-PHYS-PUB-2016-017} (referred to as the `tune').
The MC samples were produced using either the detailed ATLAS detector simulation~\cite{SOFT-2010-01} based on \textsc{Geant4}~\cite{Agostinelli:2002hh} or a faster simulation where the full \textsc{Geant4} simulation of the calorimeter response is replaced by a parameterisation of the shower shapes~\cite{SOFT-2010-01}. In both cases, the simulated events were processed through the same reconstruction software as the dataset of $pp$ collisions.
Corrections were applied to the simulated events so that the particle candidates' selection efficiencies, energy scales and energy resolutions match those determined from data control samples.
The simulated samples are normalised to their cross sections, and generated to the highest order available in perturbation theory.
 
Simulated events with pair-produced scalar LQs were generated at next-to-leading-order (NLO) in quantum chromodynamics (QCD) with \textsc{MadGraph5\_aMC@NLO} 2.6.0~\cite{Alwall:2014hca}, using the method described in Ref.~\cite{Mandal:2015lca}, in which fixed-order NLO QCD calculations~\cite{Kramer:1997hh,Kramer:2004df} are interfaced to \textsc{Pythia} 8.230~\cite{Sjostrand:2007gs} for the parton shower (PS) and hadronisation.  Parton luminosities are provided by the five-flavour scheme \verb|NNPDF3.0nlo|~\cite{Ball:2014uwa} PDF set with $\alpha_s=0.118$ and the underlying event (UE) is modelled with the A14 tune~\cite{ATL-PHYS-PUB-2014-021}. \textsc{Madspin}~\cite{Artoisenet:2013madspin} was used for the decay of the scalar \LQmixd. The QCD pair production of \LQmixd does not depend on the coupling parameter $\lambda$ to the first order, therefore the choice of this parameter is arbitrary for the production. However, the decay width depends on $\lambda$ and a narrow width is necessary for on-shell production to dominate~\cite{Buchmuller:Ruckl:Wyler,Belyaev:2005LQ}.  A small value ($\lambda=0.3$) of this parameter was set, resulting in a narrow width of about $0.2\%$ of its mass. The charge of the scalar \LQmixd is set to $-1/3e$, allowing decays into either $t\ell^{-}$ or $b\bar{\nu}$. Samples were generated for masses in the range of 1.0 TeV to 1.9 TeV in 100 GeV steps. The LQ pair-production cross sections were obtained from the calculation of  the pair production of supersymmetric partner of top quarks assuming that all other supersymmetric particles are heavier, since the production modes of this process are the same.  These cross sections are computed at approximate next-to-next-to-leading-order (NNLO) in QCD with resummation of next-to-next-to-leading logarithmic (NNLL) soft gluon terms~\cite{Beenakker:2016lwe,Beenakker:1997ut,Beenakker:2010nq,Beenakker:2016gmf}. The cross sections do not include lepton $t$-channel contributions, which are neglected in Ref.~\cite{Mandal:2015lca} and may lead to corrections at the percent level~\cite{Borschensky:2020hot}. The cross section values range from 6.8 fb to 0.02 fb for the \LQmixd mass range considered in this search.

The vector \vLQ\ signal samples were generated at leading-order (LO) in QCD with \textsc{MadGraph5$\_$aMC@NLO} 2.8.1 and the \verb|NNPDF2.3lo| PDF set using the Boston model~\cite{Diaz_2017,Bostonmodel:vLQ:1810.10017}, and matched to \textsc{Pythia} 8.244~\cite{Sjostrand:2007gs} with A14 tune. The \vLQ\ is an isospin singlet, has an electrical charge of $+5/3e$ and can couple to a top quark and a charged lepton~\cite{Buchmuller:Ruckl:Wyler}. The coupling parameter $\lambda$ was also set to 0.3. The model introduces a new massive coloured gauge boson known as heavy gluon ($g'$). There are two main coupling parameters considered for this production, labelled as $\kappa_t$ and $\kappa_s$. The $\kappa_t$ represents the coupling of the gluon to \vLQ\, while $\kappa_s$ represents the coupling of the heavy gluon $g'$ to $\tilde{U_1}$. The $\kappa_t$ = $\kappa_s$ = 0 setup corresponds to the Yang-Mills coupling scenario (\vLQy), whereas the $\kappa_t$ = $\kappa_s$ = -1 setup, referred to as the minimal coupling scenario (\vLQm), corresponds to the scenario where the Yang-Mills couplings are turned off~\cite{Diaz_2017}. The vector leptoquark samples were generated setting the $g'$ mass to zero, where each \vLQ\ decays into a top quark and a charged lepton (electron or muon), for masses in the range of 1.0 TeV to 2 TeV in steps of 100 GeV. The corresponding cross section values range from 19.7 fb to 0.02 fb (105 fb to 0.1 fb) for the minimal (Yang-Mills) coupling scenarios.

The sample used to model the \ttW (\ttdy) background process was generated using \Sherpa~2.2.10 (\Sherpa~2.2.11)~\cite{Bothmann:2019yzt}. The matrix element (ME) was calculated for up to one (zero) additional parton at NLO in QCD and up to two partons at LO in QCD using \textsc{Comix}~\cite{Gleisberg:2008fv} and \textsc{OpenLoops}~\cite{Cascioli:2011va} and merged with the \textsc{Sherpa} parton shower~\cite{Schumann:2007mg} using the \textsc{MePs@Nlo} prescription~\cite{Hoeche:2012yf}.  A CKKW merging scale of $30~\gev$ was used for the \ttW sample.
These samples are generated using the \verb|NNPDF3.0nnlo|~\cite{Ball:2014uwa} PDF set. Both the factorisation and renormalization scales are set to $\mu_R = \mu_F =
H_\textrm{T}$/2 in the \ttW sample, where $H_\textrm{T}$ is defined as the scalar sum of the transverse masses $\sqrt{p_\textrm{T}^2+m^2}$ of all final state particles.
The LO \ttW electroweak (EW) contributions are obtained from a dedicated sample simulated with \Sherpa~2.2.10 and are added together with the NLO QCD sample described above. The invariant mass of the lepton pair ($\mll$) in the \ttdy sample is set to be greater than $1~\gev$.
The complete \ttW simulation is normalised to the total cross section of $\sigma(\ttW) = 614.7$~fb that comes from the \Sherpa configuration outlined above considering NLO QCD and NLO EWK effects, based on a similar strategy as used in Ref.~\cite{Frederix:2021agh}. The \ttdy sample is normalised to the cross section $\sigma(\ttdy) = 839$~fb calculated at NLO QCD and NLO EW accuracy using \MGNLO as reported in Ref.~\cite{deFlorian:2016spz} and scaled by an off-shell correction estimated at one-loop level in \alphas.
 
Diboson ($WZ$, $ZZ$ and $WW$ including off-shell productions) background processes are simulated with \textsc{Sherpa}~2.2.2~\cite{Bothmann:2019yzt}.  The matrix element was calculated using $\textsc{Comix}$~\cite{Gleisberg:2008fv} and \textsc{OpenLoops}~\cite{Cascioli:2011va} with NLO accuracy in QCD for up to one additional parton and at LO accuracy for up to three additional partons, and merged with the \textsc{Sherpa} using \textsc{MePs@Nlo} prescription~\cite{Hoeche:2012yf}. The \verb|NNPDF3.0nnlo| set of PDFs was used, along with the dedicated set of tuned parton-shower parameters developed by the \SHERPA authors. The cross section of $\sigma(\VV) = 104$~pb was computed by \textsc{Sherpa}~2.2.2.
 
Samples for $\tth$, \ttbar, and single top production were generated using the NLO generator \textsc{Powheg-Box}~\cite{Frixione:2007nw,Nason:2004rx,Frixione:2007vw,Alioli:2010xd,Alioli:2009je,Re:2010bp} with \verb|NNPDF3.0nlo| PDF set and interfaced with \textsc{Pythia} 8 with the A14 tune. 
The $h_\text{damp}$ parameter, which controls the transverse momentum of the first additional emission beyond the Born configuration and therefore regulates the high-$\pt$ radiation, is set to $3(m_t + m_{\bar{t}} + m_{H})/4$ in the \ttH sample and to $1.5 m_t$ in the \ttbar and single top samples, where $m_t$ ($m_H$) denotes the mass of the top quark (Higgs boson).
 
A dedicated $\ttbar$ sample including relatively rare $t \to Wb \gamma^*(\to \ell^+\ell^-)$ radiative decays, $\ttbar \to W^+bW^-\bar{b}\ell^+\ell^-$, is generated using a ME calculated at LO in QCD and requiring $\mll > 1~\gev$. In this sample the photon can be radiated from the top quark, the $W$ boson, or the $b$-quark. Both the $\ttdy$ and $\ttbar \to W^+bW^-\bar{b}\ell^+\ell^-$ samples are combined and together form the ``$\ttZ$ (high mass)'' sample.
The contribution from internal photon conversions ($\gamma^* \to \ell^+\ell^-$) with $\mll < 1~\gev$ are modelled by QED multi-photon radiation via the PS in an inclusive $\ttbar$ sample and is referred to as ``$\ttbar\gamma^*$ (LM)''.
Dedicated $Z$+jets samples containing electrons from material photon conversion ($\gamma \to e^+e^-$) or internal photon conversion are generated with \textsc{Powheg-Box} and interfaced with Pythia8 for the parton showering and fragmentation. These samples are used to model the data in control regions enriched in material and internal conversion electrons, as explained in Section~\ref{sec:events}.
 
The production of \tttt events was modelled using the \textsc{Madgraph5\_aMC@NLO}~v2.6.2 generator that provides matrix elements at NLO in QCD with the \verb|NNPDF3.1nlo| PDF set matched to \textsc{Pythia}~8.186 with the A14 tune.
 
Table~\ref{tab:mcconfig} shows the configurations used in this analysis, with the samples in parentheses and in grey indicating those used to estimate the systematic uncertainties. Parton shower generator refers to the generator used for parton shower, hadronisation and underlying events. The remaining rare background contributions listed in this table are normalised using their NLO theoretical cross sections, except for the $t\bar t t$ process, for which a LO cross section is used.
 
\begin{table}[h!]
\begin{center}
\caption{\label{tab:mcconfig}
The configurations used for event generation of signal and background processes.
The samples used to estimate the systematic uncertainties are indicated in parentheses.
An electroweak boson ($W$ or $Z/\gamma^*$) is denoted as $V$ for the associated production sample with Higgs boson.
The matrix element order refers to the order in the strong coupling constant of the perturbative calculation. The ``$\ttbar W$ (EW)'' sample includes additionally next-to-leading-order electroweak corrections.
Tune refers to the underlying-event tune of the parton shower generator.
\textsc{MG5\_aMC} refers to \textsc{MadGraph5\_aMC@NLO} 2.2, 2.3, or 2.6;
\textsc{MePs@Nlo} refers to the method used in \textsc{Sherpa} to match the matrix element to the parton shower.
All samples include leading-logarithm photon emission, either modelled by the parton shower generator or by \textsc{Photos}~\cite{Golonka:2005pn}.
The mass of the top quark ($m_t$) and SM Higgs boson were set to $172.5~\gev$ and $125~\gev$, respectively.}
\vspace{0.25cm}{\small
\setlength\tabcolsep{1.5pt}
\begin{tabular}{~l^l^l^l^l^l}
\hline\hline
Process & Generator & ME order & Parton shower & PDF & Tune  \\
\hline
\LQmixd ${\mathrm{\overline{LQ}}}_{\mathrm{mix}}^{\mathrm{d}}$    & \textsc{MG5\_aMC} & NLO & \textsc{Pythia} 8.230\ & \verb|NNPDF3.0nlo| & A14~\cite{ATL-PHYS-PUB-2014-021} \\
\vLQ $\bar{\tilde{U}}_1$              & \textsc{MG5\_aMC} & LO    & \textsc{Pythia} 8.244\ & \verb|NNPDF2.3lo| & A14 \\
$\ttbar W$             & \textsc{Sherpa 2.2.10} & \textsc{MePs@Nlo} & \textsc{Sherpa} &  \verb|NNPDF3.0nnlo| & \textsc{Sherpa} default \\
\gdef\currentrowstyle{\color{gray}}
& (\textsc{MG5\_aMC})  & (NLO) & (\textsc{Pythia} 8.210) & (\verb|NNPDF3.0nlo|) & (A14)   \\
$\ttbar W$ (EW)        & \textsc{Sherpa 2.2.10} & LO & \textsc{Sherpa} &  \verb|NNPDF3.0nnlo| & \textsc{Sherpa} default \\ 
\gdef\currentrowstyle{\color{gray}}
& (\textsc{MG5\_aMC})  & (LO) & (\textsc{Pythia} 8.230) & (\verb|NNPDF3.0nlo|) & (A14)   \\
Diboson, Triboson      & \textsc{Sherpa} 2.2.2  & \textsc{MePs@Nlo} & \textsc{Sherpa} & \verb|NNPDF3.0nnlo| & \textsc{Sherpa} default  \\
$\ttbar (Z/\gamma^* \to \ell^+ \ell^-)$ & \textsc{Sherpa 2.2.11} & \textsc{MePs@Nlo} & \textsc{Sherpa} &  \verb|NNPDF3.0nnlo| & \textsc{Sherpa} default \\ 
\gdef\currentrowstyle{\color{gray}}
& (\textsc{MG5\_aMC}) & (NLO) & (\textsc{Pythia} 8.210) & (\verb|NNPDF3.0nlo|) & (A14)  \\
\ttH                   & \textsc{Powheg-BOX}  & NLO & \textsc{Pythia} 8.230\ & \verb|NNPDF3.0nlo| & A14 \\
\gdef\currentrowstyle{\color{gray}}
& (\textsc{Powheg-BOX}) & (NLO) & (\textsc{Herwig7.0.4}~\cite{Bellm:2015jjp}) &  (\verb|NNPDF3.0nlo|) &  (H7-UE-MMHT~\cite{Bellm:2015jjp}) \\
\gdef\currentrowstyle{\color{gray}}
& (\textsc{MG5\_aMC})  & (NLO) & (\textsc{Pythia} 8.230) & (\verb|NNPDF3.0nlo|) & (A14)   \\
$\ttbar \to W^+bW^-\bar{b}(\gamma* \to \ell^+ \ell^-)$ & \textsc{MG5\_aMC} & LO & \textsc{Pythia} 8.212 & \verb|NNPDF3.0lo| & A14   \\
$t (Z/\gamma^*)$       & \textsc{MG5\_aMC} & NLO & \textsc{Pythia} 8.230 & \verb|NNPDF2.3lo|  & A14  \\
$t W (Z/\gamma^*)$     & \textsc{MG5\_aMC} & NLO & \textsc{Pythia} 8.212 & \verb|NNPDF2.3lo|  & A14  \\
$\ttbar$               & \textsc{Powheg-BOX} & NLO & \textsc{Pythia} 8.230 & \verb|NNPDF3.0nlo| & A14  \\
\gdef\currentrowstyle{\color{gray}}
& (\textsc{Powheg-BOX}) & NLO & (\textsc{Herwig7.1.3}) & (\verb|NNPDF3.0nlo|) &  (H7-UE-MMHT) \\
$t\bar t t$            & \textsc{MG5\_aMC} & LO & \textsc{Pythia} 8.186 & \verb|NNPDF2.3lo| & A14 \\
$t\bar t t\bar t$      & \textsc{MG5\_aMC} & NLO & \textsc{Pythia} 8.186 & \verb|NNPDF3.1nlo| & A14 \\
Single top             & \textsc{Powheg-Box} & NLO & \textsc{Pythia} 8.230 & \verb|NNPDF3.0nlo| & A14   \\
($t$-, $Wt$-, $s$-channel) &  &  &                                     &  &  \\
$Z \to \ell^+ \ell^-$         & \textsc{Sherpa} 2.2.1 & \textsc{MePs@Nlo} & \textsc{Sherpa} & \verb|NNPDF3.0nlo| & \textsc{Sherpa} default \\ 
$Z \to \ell^+ \ell^- (\gamma \to e^{+}e^{-})$ & \textsc{Powheg-BOX} & NLO & \textsc{Pythia} 8.186 & \verb|CTEQ6L1nlo|~\cite{Pumplin:2002vw} & A14 \\
$Z \to \ell^+ \ell^- (\gamma* \to e^{+}e^{-})$ & \textsc{Powheg-BOX} & NLO & \textsc{Pythia} 8.240 & \verb|CTEQ6L1nlo| & A14 \\
$W+$jets               & \textsc{Sherpa} 2.2.1 & \textsc{MePs@Nlo} & \textsc{Sherpa} & \verb|NNPDF3.0nlo| & \textsc{Sherpa} default \\ 
$VH$                   & \textsc{Powheg-BOX} & NLO  & \textsc{Pythia} 8.186 & \verb|NNPDF3.0nlo| & A14 \\
$t\bar t W^+ W^-$      & \textsc{MG5\_aMC} & LO  & \textsc{Pythia} 8.186 & \verb|NNPDF2.3lo|  & A14  \\
$tWH$                  & \textsc{MG5\_aMC} & LO  & \textsc{Pythia} 8.235 & \verb|NNPDF2.3lo|  & A14   \\
$tHjb$                 & \textsc{MG5\_aMC} & LO  & \textsc{Pythia} 8.230 & \verb|NNPDF2.3lo|  & A14   \\
\hline\hline
\end{tabular}
}
\end{center}
\end{table}


\clearpage
\section{Event reconstruction and object identification}
\label{sec:objects}

Interaction vertices from the $pp$ collisions are reconstructed from at least two tracks with transverse momentum ($\pt$) larger than $500~\mev$ that are consistent with originating from the beam collision region in the $x$--$y$ plane. If more than one primary vertex candidate is found in the event, the candidate for which the associated tracks form the largest sum of squared $\pt$ is selected as the hard-scatter primary vertex~\cite{ATL-PHYS-PUB-2019-015}.
 
Electron candidates are reconstructed from energy clusters in the electromagnetic calorimeter matched to a track in the ID~\cite{EGAM-2018-01}. They are required to satisfy $\pt>10$~GeV and $|\eta_{\mathrm{cluster}}|<2.47$, excluding the transition region between the endcap and barrel calorimeters ($1.37<|\eta_{\mathrm{cluster}}|<1.52$).
Loose and tight electron identification working points are used~\cite{EGAM-2018-01}, based on a likelihood discriminant employing calorimeter, tracking and combined variables that provide separation between electrons and jets. The associated track of an electron candidate is required to have at least two hits in the pixel detector and seven hits total in the pixel and silicon-strip detectors combined. For the tight identification working point, one of these pixel hits must be in the innermost layer (or the next-to-innermost layer if the module traversed in the innermost layer is non-operational), and there must be no association with a vertex from a reconstructed photon conversion~\cite{PERF-2017-02} in the detector material (denoted as `material conversion' in this paper).
 
Muon candidates are reconstructed by combining tracks in the ID with tracks in the MS~\cite{MUON-2018-03}. The resulting muon candidates are re-fitted using the complete track information from both detector systems~\cite{PERF-2015-10}. They are required to satisfy $\pt > 10~\gev$ and $|\eta| < 2.5$. Loose and medium muon identification working points are used~\cite{PERF-2015-10}. Medium muon candidates with $\pt>800~\gev$ are in addition required to have hits in at least three MS stations (referred to as the `high-$\pt$ working point'), in order to maximise the momentum resolution for the muon track and thus suppress backgrounds with high-$\pt$ muons arising from momentum mismeasurements.
 
Electron (muon) candidates are matched to the primary vertex by requiring the significance of their transverse impact parameter, $d_0$, satisfies $|d_0/\sigma(d_0)|<5\,(3)$, where $\sigma(d_0)$ is the measured uncertainty in $d_0$.  Additionally, the longitudinal impact parameter, $z_0$, satisfies $|z_0 \sin\theta|<0.5$~mm, where $\theta$ is the polar angle of the track.
 
To further suppress leptons from heavy-flavour hadron decays, misidentified jets, or photon conversions (collectively referred to as `non-prompt leptons'), lepton candidates are also required to be isolated in the tracker and in the calorimeter. A track-based lepton isolation criterion is defined by calculating the quantity $I_R = \sum \pt^{\textrm{trk}}$, where the scalar sum includes all tracks (excluding the lepton candidate itself) within the cone defined by $\Delta R<R_{\textrm{cut}}$ around the direction of the lepton.  The value of $R_{\textrm{cut}}$ is the smaller of $r_{\textrm{min}}$ and $10~\gev/\pt^\ell$, where $r_{\textrm{min}}$ is set to 0.2 (0.3) for electron (muon) candidates and where $\pt^\ell$ is the lepton $\pt$. All lepton candidates must satisfy $I_R/\pt^\ell < 0.15$.
Additionally, electrons (muons) are required to satisfy a calorimeter-based isolation criterion: the sum of the transverse energy within a cone of size $\Delta R=0.2$ around the lepton, after subtracting the contributions from pile-up and the energy deposit of the lepton itself, is required to be less than 20\% (30\%) of $\pt^\ell$.
 
The selection criteria described above largely suppresses the contribution from non-prompt leptons. Additional requirements are applied in the \llSS and \lll categories to further suppress the main non-prompt lepton types. Non-prompt leptons from hadrons decays that contain bottom- and charm-quarks (denoted as `heavy-flavour (HF) non-prompt leptons') are further rejected using a boosted decision tree (BDT) discriminant, referred to as the non-prompt-lepton BDT~\cite{HIGG-2017-02}. This BDT discriminant is based on isolation and lifetime information associated with a jet reconstructed from tracks in the ID that matches the selected light lepton. Three exclusive working points (WPs) determined from the BDT score are used: \textit{Tight}, \textit{VeryTight}, and \textit{Tight-not-VeryTight}.
The \textit{Tight} WP allows to select prompt-like leptons with an efficiency for muons (barrel/endcap electrons) that satisfy the calorimeter- and track-based isolation criteria of about 60\% (60/70\%) for $\pt \sim 20~\gev$ and reaches a plateau of 95\% (95/90\%) at $\pt \sim 40\ (40/65) ~\gev$.
The prompt lepton efficiency of the \textit{VeryTight} WP for muons (barrel/endcap electrons) that satisfy the calorimeter- and track-based isolation criteria is about 55\% (55/60\%) for $\pt \sim 20~\gev$ and reaches a plateau of 90\% (85/83\%) at $\pt \sim 40\ (40/65) ~\gev$. The corresponding rejection factor\footnote{The rejection factor is defined as the reciprocal of the efficiency.} against muons (electrons) from the decay of $b$-hadrons ranges from 33 to 50 (20 to 50) for the \textit{Tight} WP, and from 50 to 100 (33 to 66) for the \textit{VeryTight} WP, depending on \pt and $\eta$, after resolving ambiguities between overlapping reconstructed objects.
The latter allows to select non-prompt-like leptons and is part of the event selection of the control regions enriched in HF non-prompt lepton background, as described in Section~\ref{sec:background}.
 
In order to further suppress electrons with incorrect charge assignment, a BDT discriminant based on calorimeter and tracking quantities~\cite{PERF-2017-01} is used.
An efficiency of approximately 96\% in the barrel region and 81\% in the endcaps is obtained, with rejection factors of 19 in the barrel region and 40 in the endcaps.
Material ($\gamma \to e^{+}e^{-}$) and internal ($\gamma^* \to e^{+}e^{-}$) conversion candidates are identified based on a combination of requirements on the invariant mass of tracks and the radius from the reconstructed displaced vertex to the primary vertex.
Material conversion candidates have a reconstructed displaced vertex with radius $r > 20$~mm that includes the track associated with the electron.\footnote{The beampipe and insertable B-layer inner radii are 23.5~mm and 33~mm, respectively.} The invariant mass of the associated track and the closest (in $\Delta\eta$) opposite-charge track reconstructed in the silicon detector, calculated at the conversion vertex, is required to be less than $100~\mev$. Internal conversion candidates, which correspond to the internal photon conversions (see Section~\ref{sec:datamc}), are required to fail the criteria for material conversions, but the invariant mass of the two tracks matched to the primary vertex is required to be less than $100~\mev$.

After the initial categorisation based on loose leptons (corresponding to $L$), the most optimal lepton working point to further optimise the event selection is chosen depending on the main background processes and statistics of each category. The medium inclusive ($M$) lepton working point corresponds to leptons passing the \textit{Tight} non-prompt-lepton BDT WP, whereas the medium exclusive ($M_{ex}$) lepton working point requires leptons to pass the \textit{Tight-not-VeryTight} non-prompt-lepton BDT WP. The tight ($T$) lepton working point selects leptons passing the \textit{VeryTight} non-prompt-lepton BDT WP and provides the highest purity of prompt leptons.
The various choices can be seen for all categories used in this analysis in Section~\ref{sec:events}.  All $M$, $M_{ex}$, and $T$ electrons in the analysis are required to not be material nor internal conversion candidates, with the exception of the electron in the control regions enriched with internal and material conversions, denoted as $e^{*}$. The various lepton working points used in this analysis are summarised in Table~\ref{tbl:tightleps}.
 
\begin{table}
\begin{center}
\caption{\label{tbl:tightleps} Description of the loose inclusive ($L$), medium inclusive ($M$), medium exclusive ($M_{ex}$), and tight ($T$)  lepton definitions. The electron $e^{*}$ is required to fulfil, in addition to the corresponding lepton definition requirements, those corresponding to an internal or material conversion candidate.}
\resizebox{\textwidth}{!}{
\begin{tabular}{l||c|c|c|c||c|c|c|c}
\hline\hline
& \multicolumn{4}{c||}{$e$} & \multicolumn{4}{c}{$\mu$} \\
\hline
Lepton definition                  & $L$ & $M$ & $M_{ex}$ & $T$ &  $L$ & $M$ & $M_{ex}$ & $T$ \\
\hline
Isolation                          &  \multicolumn{4}{c||}{Yes} & \multicolumn{4}{c}{Yes} \\
\hline
Non-prompt lepton WP               &  No & \textit{Tight} & \textit{Tight-not-}  & \textit{VeryTight} &  No & \textit{Tight} & \textit{Tight-not-}  & \textit{VeryTight}  \\
&  &   & \textit{VeryTight} & &  &  & \textit{VeryTight} &  \\
\hline
Identification                     & Loose & \multicolumn{3}{c||}{Tight} & Loose & \multicolumn{3}{c}{Medium} \\
\hline
Electron charge-misassignment veto &  No & \multicolumn{3}{c||}{Yes} & \multicolumn{4}{c}{N/A} \\
Electron conversion candidate veto &  No & \multicolumn{3}{c||}{Yes (except $e^{*}$)} & \multicolumn{4}{c}{N/A} \\
\hline
Transverse impact parameter        &  \multicolumn{4}{c||}{$<5$} & \multicolumn{4}{c}{$<3$ } \\
significance $|d_0|/\sigma_{d_0}$  & \multicolumn{4}{c||}{} &  \multicolumn{4}{c}{}  \\
\hline
Longitudinal impact parameter      &  \multicolumn{8}{c}{$< 0.5$ mm} \\
$|z_0 \sin \theta|$                &  \multicolumn{8}{c}{} \\
\hline\hline
\end{tabular}
}
\end{center}
\end{table}
 
The constituents for jet reconstruction are identified by combining measurements from both the ID and the calorimeter using a particle flow (PFlow) algorithm~\cite{PERF-2015-09,JETM-2018-05}. Jet candidates are reconstructed from these PFlow objects using the anti-$k_t$ algorithm~\cite{Cacciari:2008gp,Cacciari:2011ma} with a radius parameter of $R=0.4$. They are calibrated using simulation with corrections obtained from in situ techniques in data~\cite{JETM-2018-05}. Only jet candidates with a  $\pT> 25$~\GeV~and within $|\eta|<2.5$ are selected. In order to reduce the effect from pile-up, each jet with $\pT<60$~\GeV~and $|\eta|<2.4$ is required to satisfy the ``Tight''  working point, correspondng to an efficiency of $95\%$-$97\%$ depending on jet \pt\ , of the Jet Vertex Tagger (JVT)~\cite{ATLAS-CONF-2014-018} criteria used to identify the jets as originating from the selected primary vertex. A set of quality criteria is also applied to reject events containing at least one jet arising from non-collision sources or detector noise~\cite{ATLAS-CONF-2015-029}.
 
Jets containing $b$-hadrons are identified ($b$-tagged) via an algorithm~\cite{FTAG-2018-01,ATL-PHYS-PUB-2017-013} that uses a deep-learning neural network based on the distinctive features of the $b$-hadrons such as the impact parameters of tracks and the displaced vertices reconstructed in the ID. Additional input to this network is provided by discriminant variables constructed by a recurrent neural network~\cite{ATL-PHYS-PUB-2017-003}, which exploits the spatial and kinematic correlations between tracks originating from the same $b$-hadron. For each jet, a value for the multivariate $b$-tagging discriminant is calculated. A jet is $b$-tagged if the $b$-tagging score is above a certain threshold, referred to as an operating point (OP).
In this search, a jet is considered $b$-tagged if it passes the OP corresponding to 85\% efficiency to tag a $b$-quark jet, with a light-jet\footnote{`Light jet' refers to a jet originating from the hadronisation of a light quark ($u$, $d$, $s$) or a gluon.} rejection factor of about 40, and a charm-jet ($c$-jet) rejection factor of about 3, as determined for jets with $\pt >20~\gev$ and $|\eta|<2.5$ in simulated $\ttbar$ events.
Correction factors derived from dedicated data samples enriched in $b$-jets, $c$-jets, or light jets, are applied to the simulated samples~\cite{FTAG-2018-01,ATLAS-CONF-2018-001,ATLAS-CONF-2018-006}.
 
The ambiguities among leptons satisfying the $L$ criteria and selected jets are resolved by following an overlap removal procedure that takes into account whether a jet is $b$-tagged or not. If two electrons are closer than $\Delta R=0.1$, only the one with the higher \pt\ is considered. If an electron and a muon overlap within $\Delta R=0.1$, the muon is rejected if it is reconstructed from a track and calorimeter deposits consistent with a minimum ionising particle (i.e. calo-tagged), else the electron is rejected. If an electron and a selected jet are found within $\Delta R < 0.2$, the jet is rejected if it is not $b$-tagged~\footnote{For the overlap removal, a jet is considered $b$-tagged if it passes the 70\% working point. However, the choice of the $b$-tagging working point does not have a sizeable impact on the signal acceptance.} or if it has $\pt > 200~\gev$. Muons are required to be separated by $\Delta R > 0.4$ from any jet that is ghost-associated~\cite{Cacciari:2008gn} to it. If the jet satisfying the $\Delta R < 0.4$ requirement is not a $b$-tagged jet and contains less than three tracks with $\pt > 500~\mev$, the overlapping jet is rejected from the event, otherwise, the muon is rejected. If the overlapping jet is $b$-tagged, the muon is rejected. A lepton lying within a variable-size cone depending on the lepton \pt and with a maximum radius of $R=0.4$ around a selected jet that survived all previous overlap criteria is rejected.
 
The missing transverse momentum $\vec{p}_{\mathrm{T}}^{\mathrm{miss}}$ (with magnitude $\met$) is defined as the negative vector sum of the $\pt$ of all selected and calibrated objects in the event, including a term to account for the momentum from soft particles in the event that are not associated with any of the selected objects~\cite{PERF-2016-07}.
This soft term is calculated from inner-detector tracks matched to the selected primary vertex, which makes it more resilient to contamination from pile-up interactions.


\section{Search strategy}
\label{sec:events}

This search targets primarily scalar $\LQmixd$ pair production where each $\LQmixd$ decays into $te$ or $t\mu$. In each final state, \tete~or \tmutmu, events contain two light leptons from the $\LQmixd$ decays, plus possibly additional leptons from the top-quark(s) decay chain.  Additionally, the final state comprises two \bjets from the top-quark decays, and may contain additional light-jets from initial- or final-state radiations or from a hadronically decaying $W$ boson in one of the top-quark decays. The presence of at least three leptons in the signal-enriched event categories is exploited to substantially reduce the SM background and to improve the search sensitivity. This also makes this search orthogonal to a previous ATLAS search~\cite{EXOT-2019-19} with two oppositely charged light leptons, and thus allow for an eventual combination.
The analysis channels are sub-divided into different event categories depending on the multiplicity of light leptons and the sign of their electric charges. A maximum-likelihood fit is performed across all event categories to search for the signal and constrain several leading backgrounds simultaneously.  The kinematic reconstruction of top quarks and consequently of the LQ is difficult with multiple light leptons and neutrino(s) in the selected events.
The decay of a pair of massive LQs results in energetic final state objects. Therefore the effective mass ($\Meff$), defined as the sum of \pt of light leptons, jets, and the \met, is chosen as the final discriminant due to its power to discriminate signal against background in the search channels. The results of the search are also interpreted for the vector \vLQ\ models for each decay mode (\tete\ or \tmutmu) of the LQ pair.
 
\subsection{Event selection}
A combination of lepton triggers~\cite{TRIG-2018-01,TRIG-2018-05,TRIG-2019-04}, which are based on electron and muon signatures, are used for the events selection. Single-electron (muon) triggers are required to have an electron (muon) satisfying certain identification requirements and with a \pt above a certain threshold. For the data collected in 2015, the \pt thresholds are 24~$\gev$, 60~$\gev$ and 120~$\gev$ for the single-electron triggers and 20~$\gev$ and 50~$\gev$ for the single-muon triggers.  For the data collected during 2016-2018, the thresholds were raised slightly to 26~$\gev$, 60~$\gev$, and 140~$\gev$ for the single-electron triggers and 26~$\gev$ and 50~$\gev$ for the single-muon triggers.  The electron identification criteria are relaxed for the higher \pt threshold single-electron triggers. Single-lepton triggers with low \pt threshold and lepton isolation requirements are combined in a logical OR with the high \pt threshold triggers without isolation requirements. The dielectron trigger requires two electrons satisfying loose identification with \pt threshold of 12~$\gev$ in 2015, 17~$\gev$ in 2016, and 24~$\gev$ in 2017 and 2018. The dimuon trigger requires two muons with asymmetric \pt thresholds of 8~$\gev$ and 18~$\gev$ (8~$\gev$ and 22~$\gev$) in the 2015 (2016-2018) data period(s). The electron+muon trigger requires events to have an electron candidate satisfying loose identification with a \pt threshold of 17~$\gev$ and a muon candidate with a \pt threshold of 14~$\gev$ for all data-taking periods. Events are selected using the logical OR of the single-lepton and dilepton triggers.
 
Events selected by the trigger are required to further satisfy basic preselection requirements. They must have at least one primary vertex candidate.  Events are required to contain at least two leptons satisfying the $L$ criteria (see Section~\ref{sec:objects}) with $\pt > 10$~$\gev$. The selected light leptons are required to match, with $\Delta R < 0.15$, the corresponding leptons reconstructed by the trigger and to have a $\pt$ exceeding the trigger $\pt$ threshold by 1~$\gev$. Furthermore, two or more jets with $\pt > 25$~$\gev$ must be present in the event. 
The trigger requirement has an efficiency of 99.9\% for signal events satisfying the preselection requirements.  
 
\subsection{Event categorisation}
\label{sec:event_categ}
 
Three orthogonal final states, termed ``channels'', are analysed, defined by the multiplicity of leptons with $\pt > 10$~$\gev$:
 
\begin{itemize}
\item{\llSS: two same-charge light leptons}
\item{\lll: three light leptons}
\item{\llll: at least four leptons}
\end{itemize}
 
The channels are subdivided into different event categories optimised either to search for the signal (referred to as `signal regions', or SR), to obtain improved background estimates (referred to as `control regions', or CR), or to validate the estimated backgrounds (referred to as `validation regions', or VR). Only SRs and CRs are included in the maximum-likelihood fit.
 
Signal and control regions are defined in the \lll channels, whereas the \llll (\llSS) channel contains only signal (control) regions. Validation regions close to the signal regions are additionally present in the \lll and \llll channels.
Important background contributions to this analysis come from \ttW, \ttZ, and diboson processes. The \llSS control regions are enriched in \ttW and reducible background from non-prompt leptons, while the \lll control regions are dominated by \ttZ, diboson, and the photon conversion backgrounds.
Signal events are required to have at least three reconstructed leptons and the SRs are further optimised depending on the lepton multiplicity and kinematics of final-state objects. The signal-to-background ratio and the background compositions are different in the \lll and \llll channels. The \lll SR targets LQ pair signal events where one of the top quarks decays leptonically, whereas the \llll SR is enriched in LQ pair events where both top quarks decay into leptons.
Figure~\ref{fig:sketch_strategy} illustrates the categorisation and definition of the signal and control regions being fit simultaneously in this analysis.
 
The variable $\Meff$ is powerful in discriminating between the signal and background as shown in Figure~\ref{fig:shape_meff_tete}. An additional kinematic variable exploited in the optimisation of SRs is the invariant mass \Mllmin, defined as the minimum of all combinations of dilepton-pair invariant masses (see Figure~\ref{fig:shape_mllmin_tmtm}).  The SR labelled as SR-$e$ (SR-$\mu$) is defined for the LQ pair search in the \tete\ (\tmutmu) channel.  A total of 7 CRs are defined to provide background-rich samples that do not overlap with the SRs.  The VRs are defined, using cuts on $\Meff$ and \Mllmin, to be kinematically closer to the SRs but orthogonal to the CRs in order to validate the extrapolation of the background estimation. The definitions of all of these regions is given below.
 
In the \llSS channel, events contain two same-charge leptons with $\pt > 20$~\gev~and at least two jets, of which at least one is $b$-tagged. A CR enriched in \ttW is defined by requiring both leptons to satisfy $T$ identification criteria to suppress the contributions from non-prompt leptons. The contribution of \ttW events in this CR region is more than $50\%$ with negligible signal contamination. To estimate the non-prompt-lepton background in the \ttW CR, two additional \llSS CRs are defined. These regions are enriched in non-prompt leptons originating from heavy-flavour hadron decays. The $M_{ex}$ identification criteria is applied to at least one of the two leptons in these regions in order to be orthogonal to the signal regions. Events fulfilling the criteria $TM_{ex}$, $M_{ex}T$, or $M_{ex}M_{ex}$ for the leading and subleading leptons in $\pt$ are selected and further split according to the subleading lepton flavour. Additionally, the transverse mass of the leading lepton and the missing transverse energy, $m_T$($\ell_0$, \met), defined as $\sqrt{2 \met p_{T, \ell_0} (1- \cos(\phi_{\ensuremath{miss}}-\phi_{\ell_0}))}$, is required to be lower than 250~$\GeV$ in the $TM_{ex}$ and $M_{ex}T$ regions, in order to reduce the \ttW contribution in these CRs. The region where the subleading lepton is an electron (muon) is labelled as \lltte (\llttmu). The selections applied to these three CRs are given in Table~\ref{tab:llSSCRs}.
 
In the \lll channel, events are required to contain three leptons, where two same-charge leptons satisfy the $M$ identification, and the remaining lepton passes the $L$ identification. Furthermore, events contain one opposite-charge (OS) and same-flavour (SF) pair having invariant mass, $\mllOSSF > 12$~\GeV. Selected events fall into seven event categories: two SRs, four CRs and one VR. Regions enriched in diboson and \ttZ background are defined by requiring $\mllOSSF$ satisfying criteria compatible with a $Z$ boson, $|\mllOSSF - m_Z| < 10$~\GeV, differing in the $b$-jet multiplicity requirement. These regions are labelled as \lllVV and \lllttZ for the diboson and \ttZ CRs, respectively.  Both CRs have respective event purities of more than $60\%$. Two additional control regions enriched in photon conversions from $Z \rightarrow \mu \mu \gamma^*(\rightarrow ee)$ are defined, according to the identification of the electron as a material conversion or internal conversion candidate, and are labelled as \lllmatc and \lllintc, respectively.  These regions are also required to have invariant mass of three leptons, $\mlll$, compatible with a $Z$ boson, $|\mlll - m_{Z}|< 10$~\GeV, while satisfying $|\mllOSSF - m_Z| > 10$~\GeV.  Events consisting of three leptons with $\pt > 20$~\GeV, at least two jets, and at least one $b$-tagged jet are split into two SRs depending on the multiplicity of leptons with certain flavour. The \lll SR-$e$ (SR-$\mu$) is characterised by at least two electrons (muons), and additionally has $\Mllmin \geq 200$~\GeV\ and $\Meff \geq 500$~\GeV. One VR is defined by removing the \Meff requirement and inverting the \Mllmin cut from the SR$-e$ and SR-$\mu$ criteria and merging the electron and muon channels. Selection criteria of the \lll CRs, VR and SRs are summarised in Table~\ref{tab:3lCRsSRs}.
 
In the \llll channel, events are required to have four $L$ identified leptons, at least two jets, and at least one $b$-tagged jet. Events with an OS and SF lepton pair compatible with a $Z$ boson are vetoed.  Selected events are split into two SRs and one VR. The $\llll$SR-$e$ and $\llll$SR-$\mu$ are characterised by $\Mllmin \geq 100$~\GeV\ and $\Meff \geq 500$~\GeV, while $\llll$VR has $\Mllmin < 100$~\GeV, no requirement on \Meff, and has the electron and muon regions merged. Table~\ref{tab:4lSRs} summarises the definitions of the SRs and VR in the \llll channel.
 
The total signal acceptance $\times$ efficiency in the \tete\ (\tmtm) selections ranges from $13.9\%$ to $15.4\%$ ($13.3\%$ to $17.5\%$) depending on the \LQmixd mass. Similar efficiencies are obtained for the $\tilde{U}_{1}$ signal.
 
\begin{figure}[h!]
\centering
\includegraphics[width=1.1\textwidth]{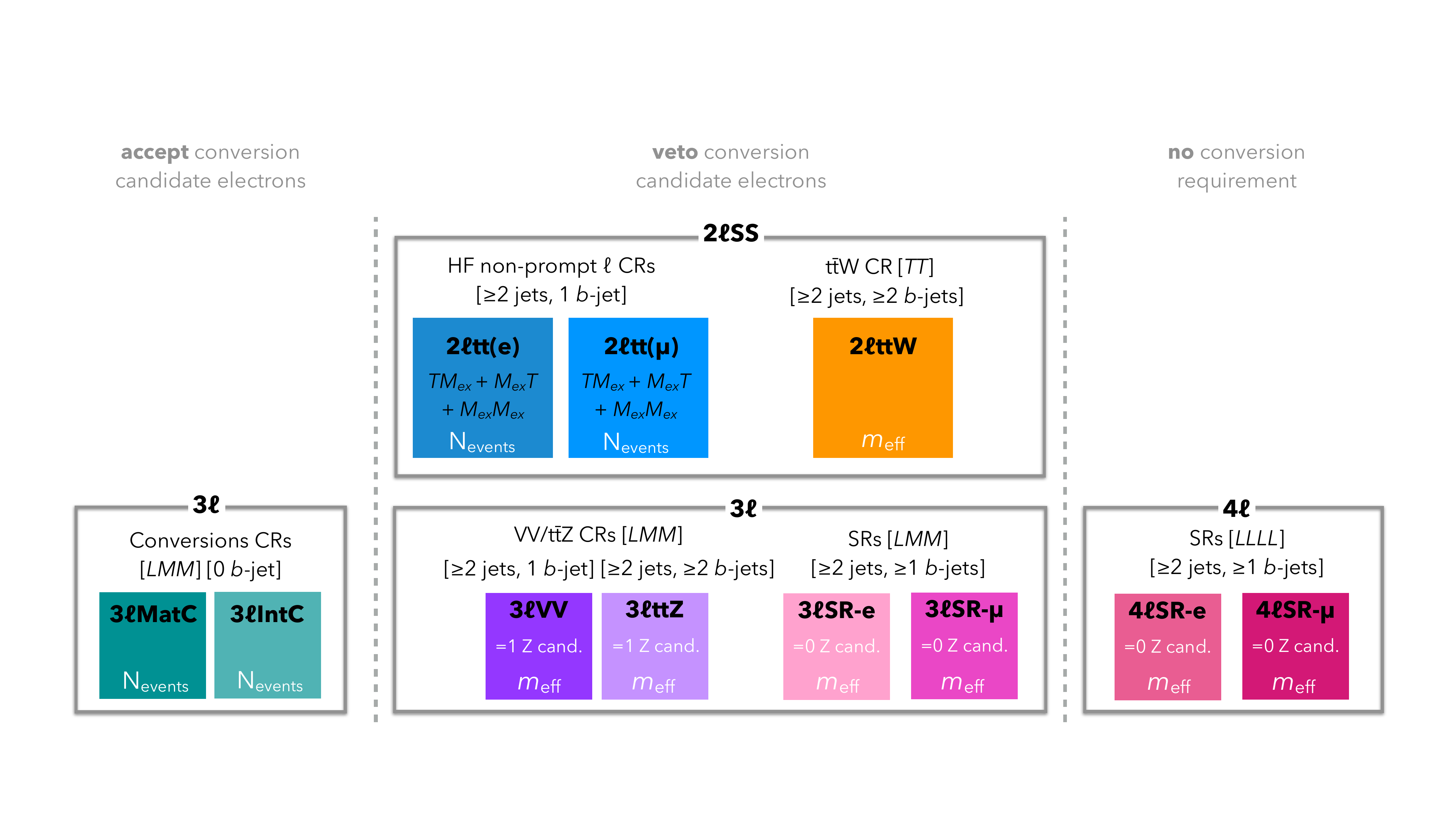}\\
\caption{Illustrative sketch of the definition of the signal and control regions. The corresponding observable used in the simultaneous fit, as described in Section~\ref{sec:results}, is given at the bottom of each region box.}
\label{fig:sketch_strategy}
\end{figure}
 
\begin{figure*}[htb]
\centering
\subfloat[]{\includegraphics[width=0.45\textwidth]{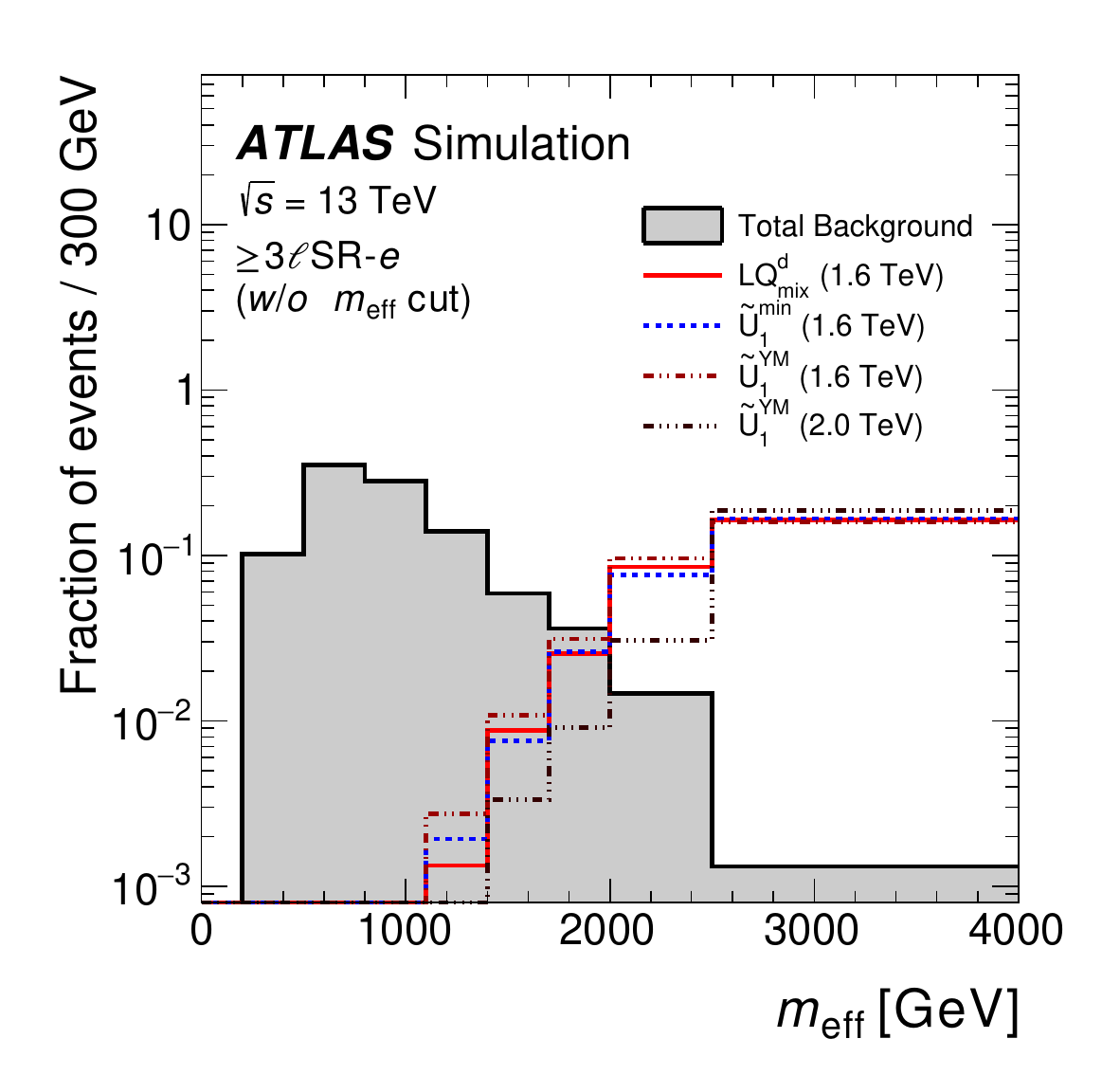}\label{fig:shape_meff_tete}}
\subfloat[]{\includegraphics[width=0.45\textwidth]{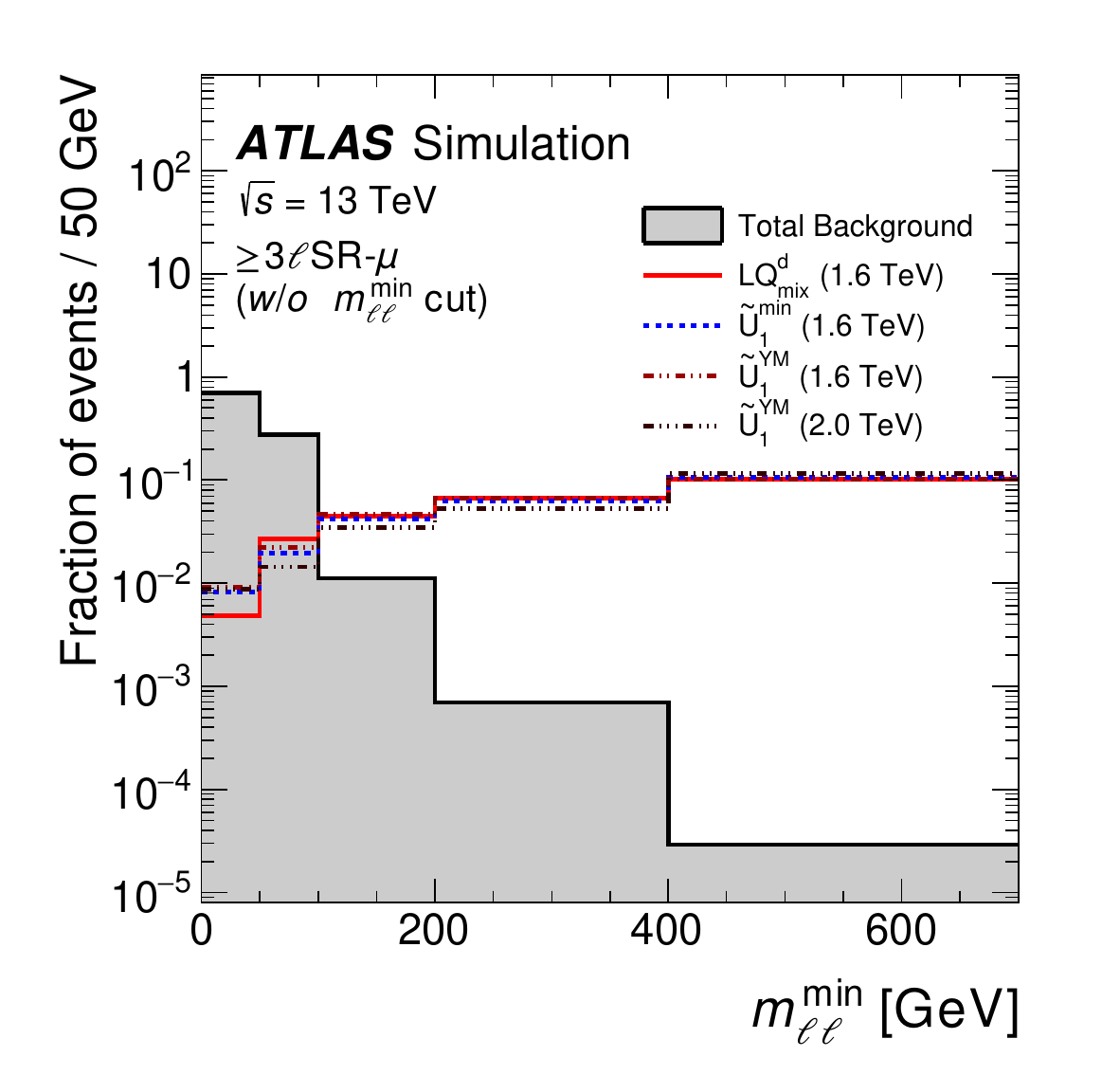}\label{fig:shape_mllmin_tmtm}}
\caption{Comparison of the distribution of (a) $\Meff$ in $\lll$SR-$e$ or $\llll$SR-$e$ region without an $\Meff$ requirement, and (b) \Mllmin in $\lll$SR-$\mu$ or $\llll$SR-$\mu$ region without a \Mllmin requirement, between the total background (shaded histogram) and the various LQ signals.
The shapes of these variables ($\Meff$ and \Mllmin) are very similar in the electron and muon channels.
The last bin in each distribution contains the overflow.
}
\end{figure*}
 
\begin{table*}[htb]
\centering
\caption{\label{tab:llSSCRs}
Summary of event categories for the \llSS control regions. All events are required to satisfy preselection requirements. The exclusive medium lepton and tight lepton definitions (see Table~\ref{tbl:tightleps}) are denoted as $M_{ex}$ and $T$, respectively. The lepton ordering of the ${e/\mu}$ selection and combination rows is based on the lepton $\pt$, with the first one corresponding to the leading lepton. The $e$ internal (material) conversion veto refers to the requirement where the electron candidate fails to fulfil the criteria for the internal (material) photon conversion (see Section~\ref{sec:objects}). Dashes (---) refer to cuts that are not applicable.}
\vspace*{0.25cm}
\small
\def\arraystretch{1.4}
\begin{tabular}{l|cc|c}
\hline\hline
& \multicolumn{3}{c}{\llSS CRs} \\
& \lltte & \llttmu & \llttW \\
\hline\hline
${e/\mu}$ selection & \multicolumn{2}{c|}{{\small$TM_{ex}$ || $M_{ex}T$ || $M_{ex}M_{ex}$}} & $TT$ \\
${e/\mu}$ combination & $ee$/$\mu e$ & $\mu\mu$/$e\mu$ & $ee$/$\mu\mu$/$e\mu$/$\mu e$  \\
$\ell \ell$ charge & \multicolumn{3}{c}{$+ +$ or $- -$} \\
$e$ internal conversion veto & \multicolumn{3}{c}{Yes} \\
$e$ material conversion veto & \multicolumn{3}{c}{Yes} \\ \hline
Number of jets & \multicolumn{3}{c}{$\geq 2$} \\
Number of $\mathrm{b\mathchar`-jets}$ & \multicolumn{2}{c|}{$1$} & $\geq 2$ \\ \hline
$p_{\mathrm T}^{\ell}$ [\gev] & \multicolumn{3}{c}{$> 20$}  \\ \hline
$m_{\mathrm T}(\ell_0, \met)$ [\gev] & \multicolumn{2}{c|}{$<$ 250} & --- \\
\hline\hline
\end{tabular}
\end{table*}
 
\begin{table*}[htb]
\centering
\caption{\label{tab:3lCRsSRs}
Summary of control regions and signal regions categories in the \lll channel. All events are required to satisfy the preselection requirements. The medium and loose lepton definitions (see Table~\ref{tbl:tightleps}) are labelled as $M$ and $L$, respectively. Same-charge (opposite-charge) lepton pairs are also referred to as same-sign (opposite-sign) with abbreviation SS (OS), and the invariant mass of two same flavour and oppositely charged lepton pair is labelled as \mllOSSF. The lepton with opposite-sign charge to that of the other two same-sign leptons is denoted $\ell_{0}$, but is not necessarily the one with highest $\pt$; the remaining SS leptons are denoted $\ell_{1}$ and $\ell_{2}$. The $e$ internal (material) conversion veto refers to the requirement where the electron candidate fails to fulfil the criteria for the internal (material) photon conversion (see Section~\ref{sec:objects}). No validation regions are included in the fit. Dashes (---) refer to cuts that are not applicable.}
\vspace*{0.25cm}
\small
\resizebox{\textwidth}{!}{
\def\arraystretch{1.4}
\begin{tabular}{l|cccc|c|cc}
\hline\hline
& \multicolumn{7}{c}{\lll} \\  \hline
& \multicolumn{4}{c|}{CR} & VR & \multicolumn{2}{c}{SR}  \\
& \lllVV & \lllttZ & \lllintc & \lllmatc & $\lll$VR  & $\lll$SR-$e$ & $\lll$SR-$\mu$  \\ \hline\hline
${e/\mu}$ selection & \multicolumn{7}{c}{$M$ (SS pair), $L$ other}  \\
${e/\mu}$ combination & \multicolumn{5}{c|}{$3e$ / $2e1\mu $ / $2\mu 1e$ / $3\mu$} & \multicolumn{1}{c|}{$3e$ / $2e1\mu $} & \multicolumn{1}{c}{$3\mu$ / $2\mu 1e$} \\
Total charge & \multicolumn{2}{c|}{$\pm 1$} & \multicolumn{2}{c|}{---} & \multicolumn{3}{c}{$\pm 1$}\\
$e$ internal conversion & \multicolumn{2}{c|}{Yes} & Inverted & Yes & \multicolumn{3}{c}{Yes} \\
veto                    &  \multicolumn{2}{c|}{}  & ($\ell_{1}$ or $\ell_{2}$) & ($\ell_{1}$ and $\ell_{2}$) & \multicolumn{3}{c}{} \\
$e$ material conversion & \multicolumn{2}{c|}{Yes} & Yes & Inverted & \multicolumn{3}{c}{Yes} \\
veto                    &  \multicolumn{2}{c|}{}  & ($\ell_{1}$ and $\ell_{2}$) & ($\ell_{1}$ or $\ell_{2}$) &  \multicolumn{3}{c}{} \\ \hline
Number of jets & \multicolumn{2}{c|}{$\geq 2$} & \multicolumn{2}{c|}{$\geq 0$} & \multicolumn{3}{c}{$\geq 2$} \\
Number of $\mathrm{b\mathchar`-jets}$ & $1$ & \multicolumn{1}{c|}{$\geq 2$} & \multicolumn{2}{c|}{$0$} & \multicolumn{3}{c}{$\geq 1$} \\ \hline
$p_{\mathrm T}^{\ell}$ [\gev] & \multicolumn{4}{c|}{$> 20$ (SS pair), $>10$ other} & \multicolumn{3}{c}{$>20$} \\  \hline
\mllOSSF[\gev] & \multicolumn{7}{c}{$> 12$} \\
$\lvert \mllOSSF - m_{Z} \lvert$ [\gev] & \multicolumn{2}{c|}{$< 10$} & \multicolumn{2}{c|}{$> 10$} & \multicolumn{3}{c}{$> 10$} \\
$\lvert \mlll - m_{Z} \lvert$ [\gev] & \multicolumn{2}{c|}{---} & \multicolumn{2}{c|}{$< 10$} & \multicolumn{3}{c}{---} \\ \hline
\Mllmin[\gev] & \multicolumn{4}{c|}{---} & $< 200$ & \multicolumn{2}{c}{$\geq 200$}  \\ \hline
\Meff[\gev] & \multicolumn{4}{c|}{---} & --- & \multicolumn{2}{c}{$\geq 500$}   \\
\hline\hline
\end{tabular}
}
\end{table*}
 
\begin{table*}[htb]
\centering
\caption{\label{tab:4lSRs}
Summary of signal regions categories in the \llll channel. All events are required to satisfy the preselection requirements. The loose lepton definition (see Table~\ref{tbl:tightleps}) is labelled as $L$. In \llll channel the leptons are ordered in \pt. The signal region consisting of two electrons and two muons is associated to $\llll$SR-$e$\ ($\llll$SR-$\mu$), if the leading light lepton in \pt is an electron (a muon). No validation regions are included in the fit. Dashes (---) refer to cuts that are not applicable.}
\vspace*{0.25cm}
\small
\def\arraystretch{1.4}
\begin{tabular}{l|c|cc}
\hline\hline
& \multicolumn{3}{c}{\llll} \\  \hline
& VR & \multicolumn{2}{c}{SR} \\
& $\llll$VR & $\llll$SR-$e$ & $\llll$SR-$\mu$ \\
\hline\hline
${e/\mu}$ selection & \multicolumn{3}{c}{$L$}  \\
${e/\mu}$ combination & $4e$ / $3e 1\mu$ / $2e 2\mu$ / $3\mu 1e$ / $4\mu$  & \multicolumn{1}{c|}{$4e$ / $3e 1\mu$ / $2e 2\mu$ (lead $e$)} & \multicolumn{1}{c}{$4\mu$ / $3\mu 1e$ / $2\mu 2e$ (lead $\mu$)} \\
Total charge & \multicolumn{3}{c}{0} \\ \hline
Number of jets & \multicolumn{3}{c}{$\geq 2$} \\
Number of $\mathrm{b\mathchar`-jets}$ & \multicolumn{3}{c}{$\geq 1$} \\ \hline
$p_{\mathrm T}^{\ell}$ [\gev] & \multicolumn{3}{c}{$>10$} \\  \hline
\mllOSSF[\gev] & \multicolumn{3}{c}{$> 12$} \\
$\lvert \mllOSSF - m_{Z} \lvert$ [\gev] & \multicolumn{3}{c}{$> 10$} \\ \hline
\Mllmin[\gev] & $< 100$ & \multicolumn{2}{c}{$\geq 100$} \\ \hline
\Meff[\gev] & --- & \multicolumn{2}{c}{$\geq 500$} \\
\hline\hline
\end{tabular}
\end{table*}


\clearpage
\section{Background estimation}
\label{sec:background}

The background processes passing the signal region selections are categorised into irreducible and reducible backgrounds.
Irreducible backgrounds  (Section~\ref{sec:bkg_irreducible}) have only prompt leptons, i.e.\ produced in $W/Z$ boson decays, in leptonic $\tau$-lepton decays, or internal conversions. Reducible backgrounds (Section~\ref{sec:bkg_reducible}) have prompt leptons with misassigned charge or at least one non-prompt lepton.
 
All backgrounds with the exception of electrons with misassigned charge (denoted as QMisID) are estimated using the simulated samples described in Section~\ref{sec:datamc}. In some cases, the simulation is improved using additional corrections derived in data control samples before the simultaneous likelihood fit to data. In particular, the $\njets$ and $\nbjets$ of the simulated diboson and non-prompt-lepton backgrounds, respectively, require dedicated corrections, whereas the normalisation of the background contribution from simulated \ttbar events with at least one additional \bjet (\ttbar+$\geq 1$ \bjet) is scaled to better describe the data. In addition, the yields of some simulated backgrounds, in particular $\ttW$, $\ttZ$, diboson and non-prompt-lepton backgrounds, are adjusted via normalisation factors that are determined by performing the likelihood fit to data across all event categories (control and signal regions as defined in Tables~\ref{tab:llSSCRs},~\ref{tab:3lCRsSRs} and~\ref{tab:4lSRs}), discussed in Section~\ref{sec:results}.
 
\subsection{Irreducible backgrounds}
\label{sec:bkg_irreducible}
 
Background contributions with prompt leptons originate from a wide range of physics processes with the relative importance of individual processes varying by channel. The main irreducible backgrounds originate from $\ttW$, $\ttdy$, and diboson (in particular $WZ$) production, and have final states and kinematic properties similar to the leptoquark signal. Smaller contributions originate from the following rare processes: $\tZ$, $\tW$, $\tWZ$, $\ttWW$, triboson, $\ttt$ and $\tttt$ production.
 
\subsubsection{\ttW background}
 
The $\ttW$ background represents the leading background in several event categories. Despite the use of state-of-the-art simulations, accurate modelling of additional QCD and QED radiation in $\ttW$ production remains challenging. Since the signal leptoquark events populate mostly the \lll and \llll regions, the \llSS region is split based on various combinations of lepton requirements (see Table~\ref{tbl:tightleps}) and $b$-jet multiplicities in order to create control regions enriched in non-prompt leptons or \ttW. The \llttW control region is required to have exactly two same-sign $T$ leptons and $\geq$ 2 b-jets. The \Meff distribution is the variable fitted in this category and provides additional constrain on this background in the tails of the distribution.
 
An overall normalisation factor is derived from the final simultaneous likelihood background-only fit that compares the data-MC agreement in this region. The measured normalisation factor is $\hat{\lambda}_{\ttW} = 1.27 \pm 0.15$, which is compatible with that determined in the ATLAS SM $t\bar{t}t\bar{t}$ analysis~\cite{Aad:2020klt}, and with a previous ATLAS measurement of the \ttW production cross section~\cite{Aaboud:2019njj}.
 
Figure~\ref{fig:ttWCR} displays the $\Meff$ distribution in the \llttW CR after the likelihood fit to data. There is good agreement between data and the background prediction after the fit within the statistical and systematic uncertainty band at high values of the $\Meff$ distribution.
 
\begin{figure}[h!]
\centering
\subfloat[]{\includegraphics[width=0.45\textwidth]{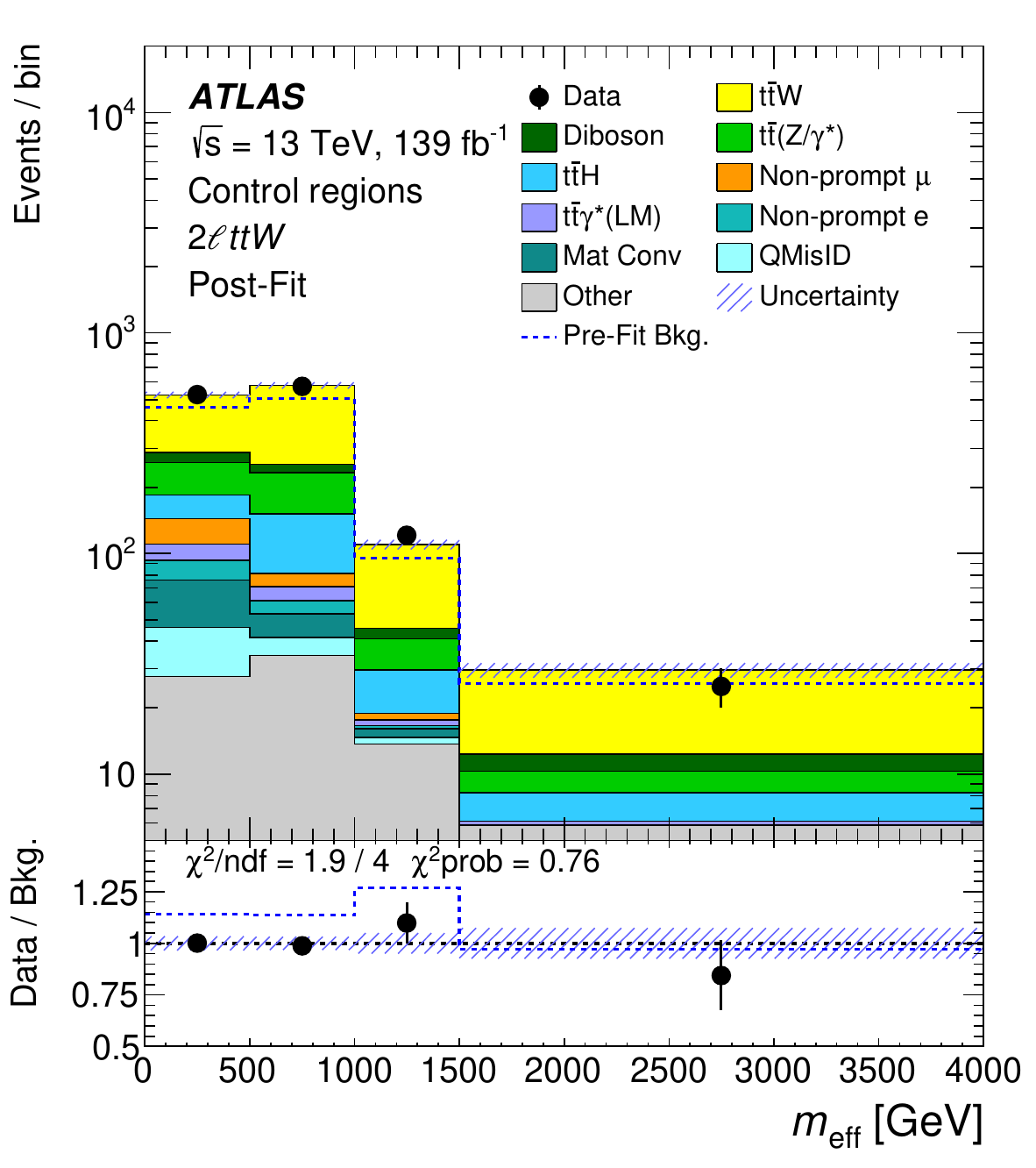}\label{fig:ttWCR}}
\subfloat[]{\includegraphics[width=0.45\textwidth]{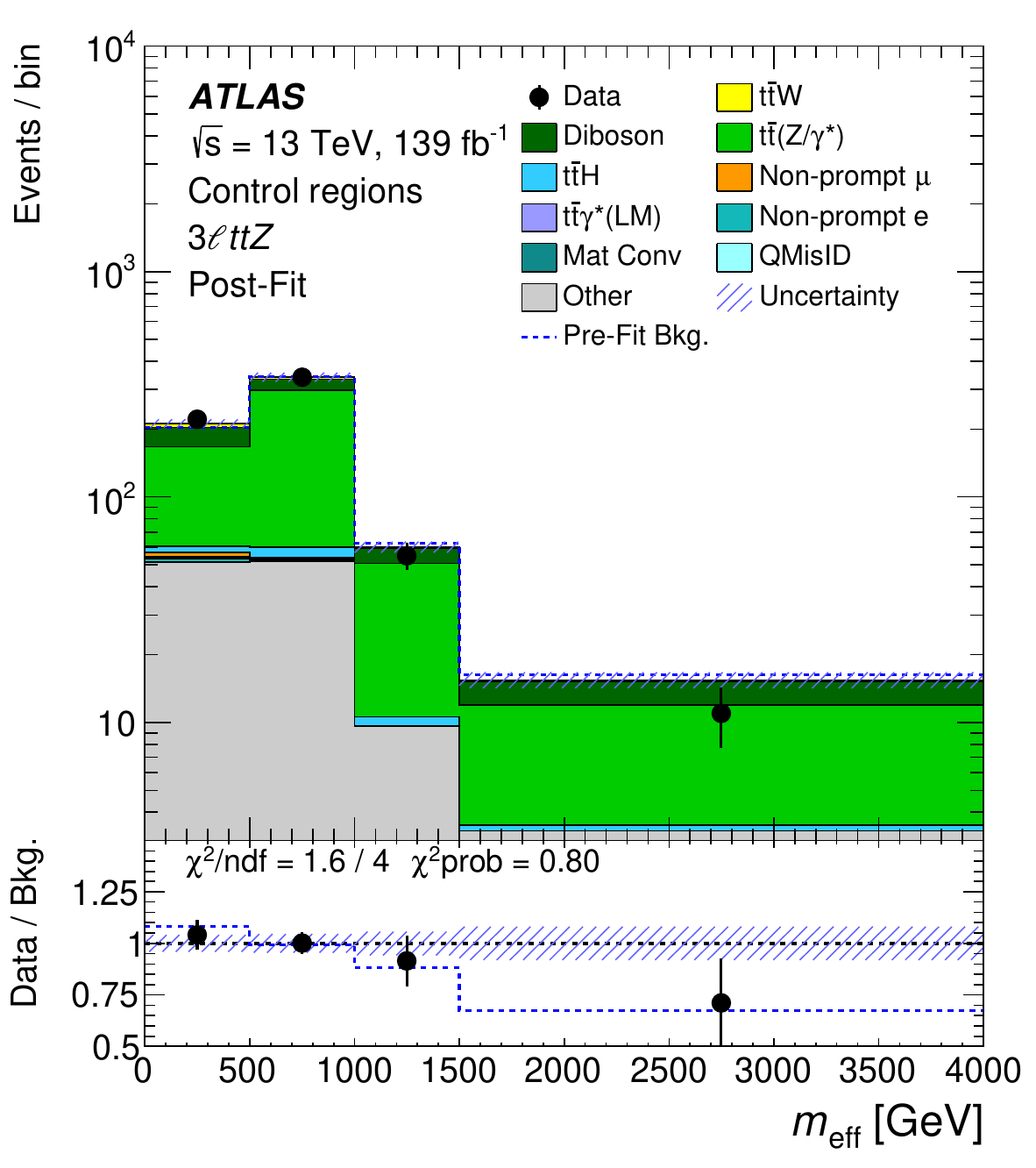}\label{fig:ttZCR}}\\
\subfloat[]{\includegraphics[width=0.45\textwidth]{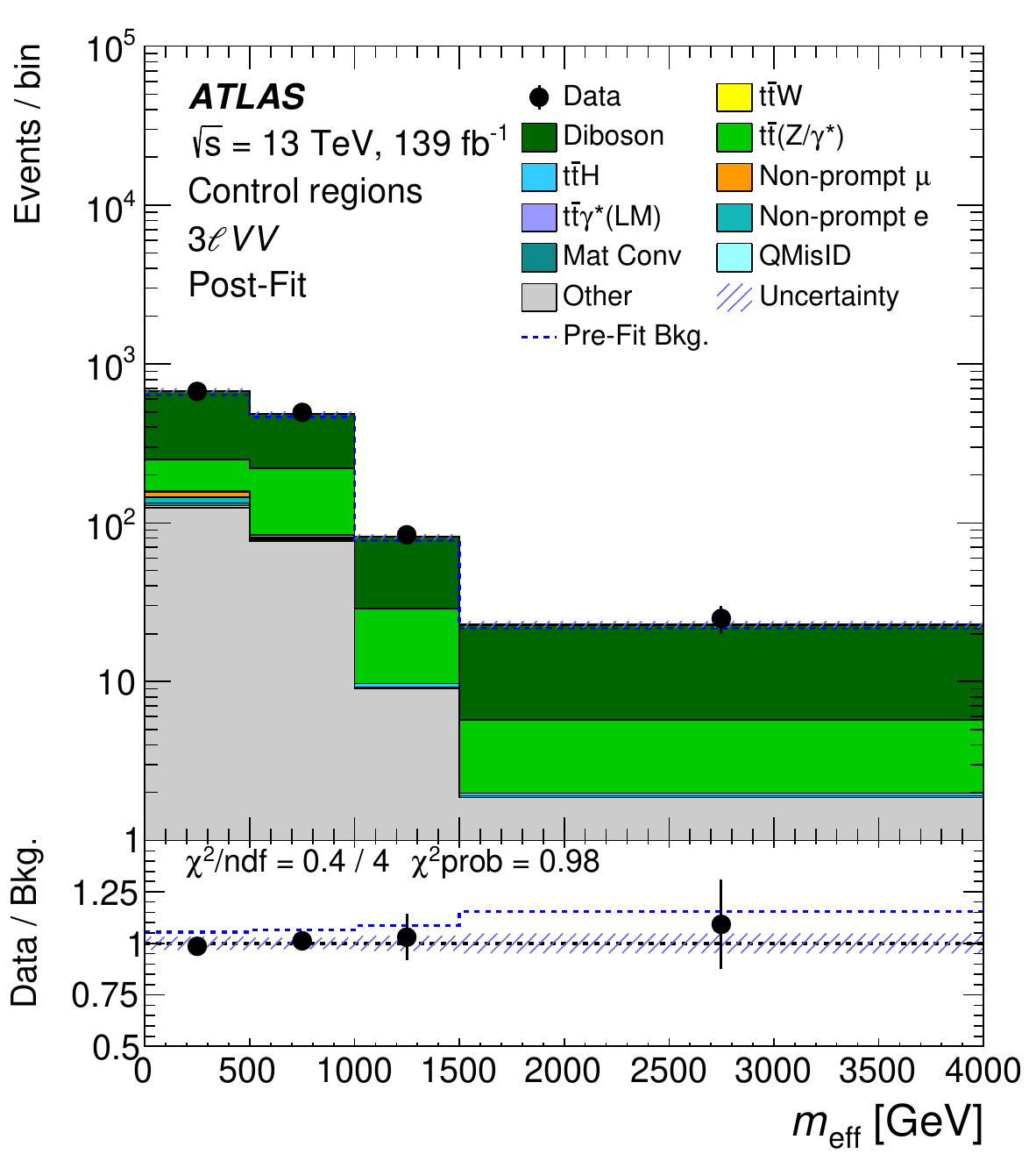}\label{fig:VVCR}}
\caption{
Comparison between data and the background prediction (``Bkg.'') for the \Meff distribution in the (a) \llttW, (b) \lllttZ, and (c) \lllVV CRs.
The background contributions after the likelihood fit to data (``Post-Fit'') under the background-only hypothesis are shown as filled histograms.
The total background prediction before the likelihood fit to data (``Pre-Fit Bkg.'') is shown as a dashed blue histogram in the upper panel.
The ratio of the data to the background prediction is shown in the lower panel, separately for post-fit background (black points) and pre-fit background (dashed blue line).
The size of the combined statistical and systematic uncertainty in the background prediction is indicated by the blue hatched band.
The last bin in each figure contains the overflow.}
\end{figure}
 
\subsubsection{Diboson and \ttdy backgrounds}
 
The diboson simulated sample does not correctly model the jet multiplicity spectrum in data. Therefore, a data-driven correction is derived from an inclusive $\lll$ diboson-enriched region with zero $b$-jets and at least one jet, referred to as \lllVVnb region. The events are required to have three leptons passing the ($L$, $M$, $M$) selection, where the second and third are the same-sign lepton pair, and the first lepton is opposite-sign with respect to the other two. The $\njets$ correction values range between 0.99 for diboson events with one additional jet to 0.67 for diboson events with at least seven additional jets.
 
The \lllVV and \lllttZ CRs are used in the likelihood fit to improve the prediction of the diboson background contribution with at least one $b$- or $c$-jet (denoted as $\VV+\mathrm{HF}$) and $\ttdy$ processes, respectively. The number of $b$-jets provides good discrimination between these two processes and is used to build the control regions. The discriminating variable used in the fit is the $\Meff$ distribution. The measured normalisation factors from the final simultaneous likelihood background-only fit are: $\hat{\lambda}_{\VV+\mathrm{HF}} = 1.21 \pm 0.18$ and $\hat{\lambda}_\textrm{\ttZ} = 0.98 \pm 0.09$.
 
Figures~\ref{fig:ttZCR} and ~\ref{fig:VVCR} display the $\Meff$ distributions in the \lllVV and \lllttZ CRs after the likelihood fit to data. There is good agreement between data and the background prediction after the fit within the statistical and systematic uncertainty band at high values of the $\Meff$ distribution.

\subsubsection{Other irreducible backgrounds}
 
The rate of the background from internal conversions with $m(e^+e^-)<1~\gev$ is estimated using the two dedicated CRs (\lllintc and \lllmatc), where at least one internal or material conversion electron candidate is required, respectively. The total yield in each category is used in the final likelihood background-only fit to determine the following normalisation factor: $\hat{\lambda}^\textrm{IntC}_e = 1.09 \pm 0.24$, where the uncertainty is dominated by the statistical uncertainty.

\subsection{Reducible backgrounds}
\label{sec:bkg_reducible}
 
\subsubsection{Non-prompt leptons}
 
Non-prompt leptons originate from material conversions, heavy-flavour hadron decays, or the improper reconstruction of other particles, with an admixture strongly depending on the lepton quality requirements and varying across event categories. These backgrounds are in general small in the \lll SR and are estimated from simulation, with the normalisation determined by the likelihood fit. The non-prompt lepton background contribution in the \llll SR is negligible and is therefore taken from simulation without dedicated data-driven corrections. The main contribution to the non-prompt-lepton background is from $\ttbar$ production, followed by much smaller contributions from $V$+jets and single-top-quark processes. The non-prompt leptons in the simulated samples are labelled according to whether they originate from HF or light-flavour (LF) hadron decays, or from a material conversion candidate. The HF category includes leptons from both bottom and charm decays.
 
Two corrections are applied to the \ttbar and the overall non-prompt lepton background simulation before the fit. First, the \ttbar+$\geq 1$ \bjet contribution is corrected by a factor of 1.3 as measured by a previous ATLAS analysis sensitive to the in-situ measurement of this contribution in the single- and opposite-sign di-lepton final states~\cite{TOPQ-2020-10}.
Second, the shape of the $b$-jet multiplicity in the non-prompt lepton background simulation is corrected to match data in an orthogonal \llSS validation region enriched with non-prompt leptons, where one of the leptons in required to pass a looser non-prompt lepton BDT score but not pass the $M$ lepton WP.
 
Several of the event categories introduced in Section~\ref{sec:events} were designed to be enriched in specific processes and are used to derive normalisation factors to improve their modelling by the simulation.
In three control regions (\lllmatc, \lltte, \llttmu) enriched in \ttbar events with non-prompt leptons, the total event yield is used in the likelihood fit to estimate the normalisation factors for these three non-prompt-lepton background contributions. The estimate of non-prompt leptons from heavy-flavour hadron decays or the improper reconstruction of other particles is measured in the \lltte and \llttmu control regions within the simultaneous fit and only affects the same-sign pair in \llSS and \lll. The normalisation factor for HF non-prompt leptons is measured separately for electrons and muons, $\lambda^\textrm{had}_e$ and $\lambda^\textrm{had}_\mu$ respectively. An additional normalisation factor is determined for the material conversion background, $\lambda^\textrm{MatC}_e$. The measured normalisation factors from the final simultaneous likelihood background-only fit are: $\hat{\lambda}^\textrm{had}_e = 0.86 \pm 0.35$,  $\hat{\lambda}^\textrm{had}_\mu = 1.05 \pm 0.22$, and $\hat{\lambda}^\textrm{MatC}_e = 1.26 \pm 0.39$, where the uncertainties are dominated by the statistical uncertainty.

\subsubsection{Charge misassignment}
 
The background process containing electrons with the charge incorrectly assigned affects primarily the $\ll$ channel. This background predominantly arises from $t\bar{t}$ production, with one electron having undergone a hard bremsstrahlung as well as an asymmetric conversion ($e^\pm \to e^\pm \gamma^* \to e^\pm e^+ e^-$) or a mismeasured track curvature.
The muon charge misassignment rate is negligible in the $\pt$ range relevant to this analysis. The electron charge misassignment rate is measured in data using samples of
$Z \to e^+e^-$ events reconstructed as same-charge pairs and as opposite-charge pairs.  The background in these events is estimated from the sideband of $m_{e^+ e^-}$ distribution and subtracted from data.
 
The charge misassignment rate is parameterised as a function of electron $\pt$ and $|\eta|$. It varies from about $10^{-5}$ for low-$\pt$ electrons ($17 \leq \pt \leq 50~\gev$) at $|\eta| \leq 1.37$, to about $3 \times 10^{-4}$ for high-$\pt$ electrons ($\pt \geq 100~\gev$) at $1.52 \leq  |\eta| \leq 2$.
The measured charge misassignment rate is then applied to data events satisfying the requirements of the $\ll$ channels, except that the two leptons are required to be of opposite charge, to estimate the QMisID background in each of the corresponding event categories.
 
\subsection{Validation of background modelling}
 
Two validation regions are defined by requiring lower values of \Mllmin than in the signal regions, as described in Section~\ref{sec:event_categ}. Figure~\ref{fig:meffspectrum_VRs} shows the $\Meff$ distribution in the $\lll$VR and $\llll$VR regions after the likelihood fit to data. The $\lll$VR is composed of 30\% \ttW production, followed by \ttZ and diboson production, whereas the $\llll$VR is enriched in \ttZ and diboson processes. There is good agreement between data and the background prediction after the fit within the statistical and systematic uncertainty band.
 
\begin{figure*}[h!]
\centering
\subfloat[]{\includegraphics[width=0.45\textwidth]{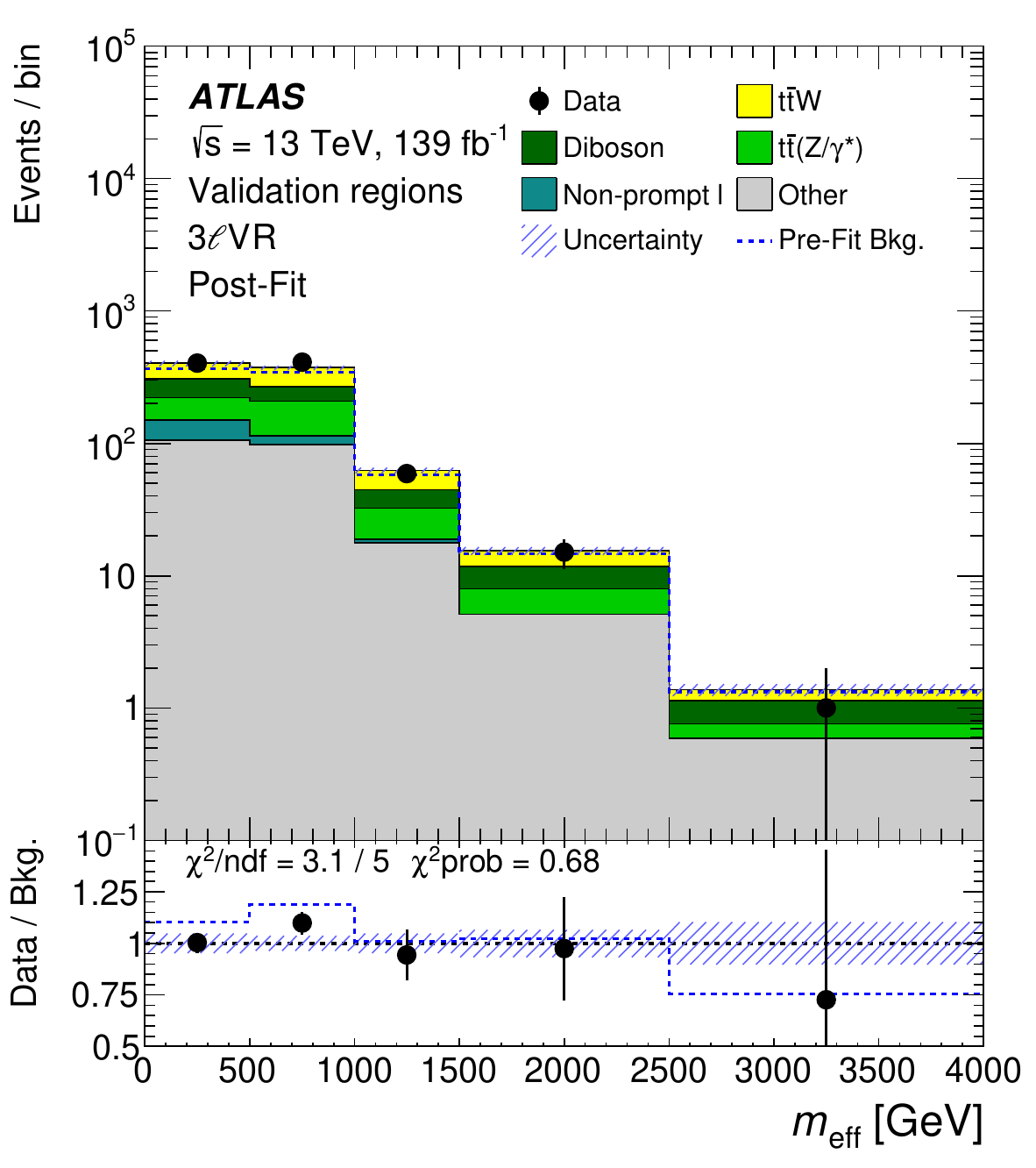}}
\subfloat[]{\includegraphics[width=0.45\textwidth]{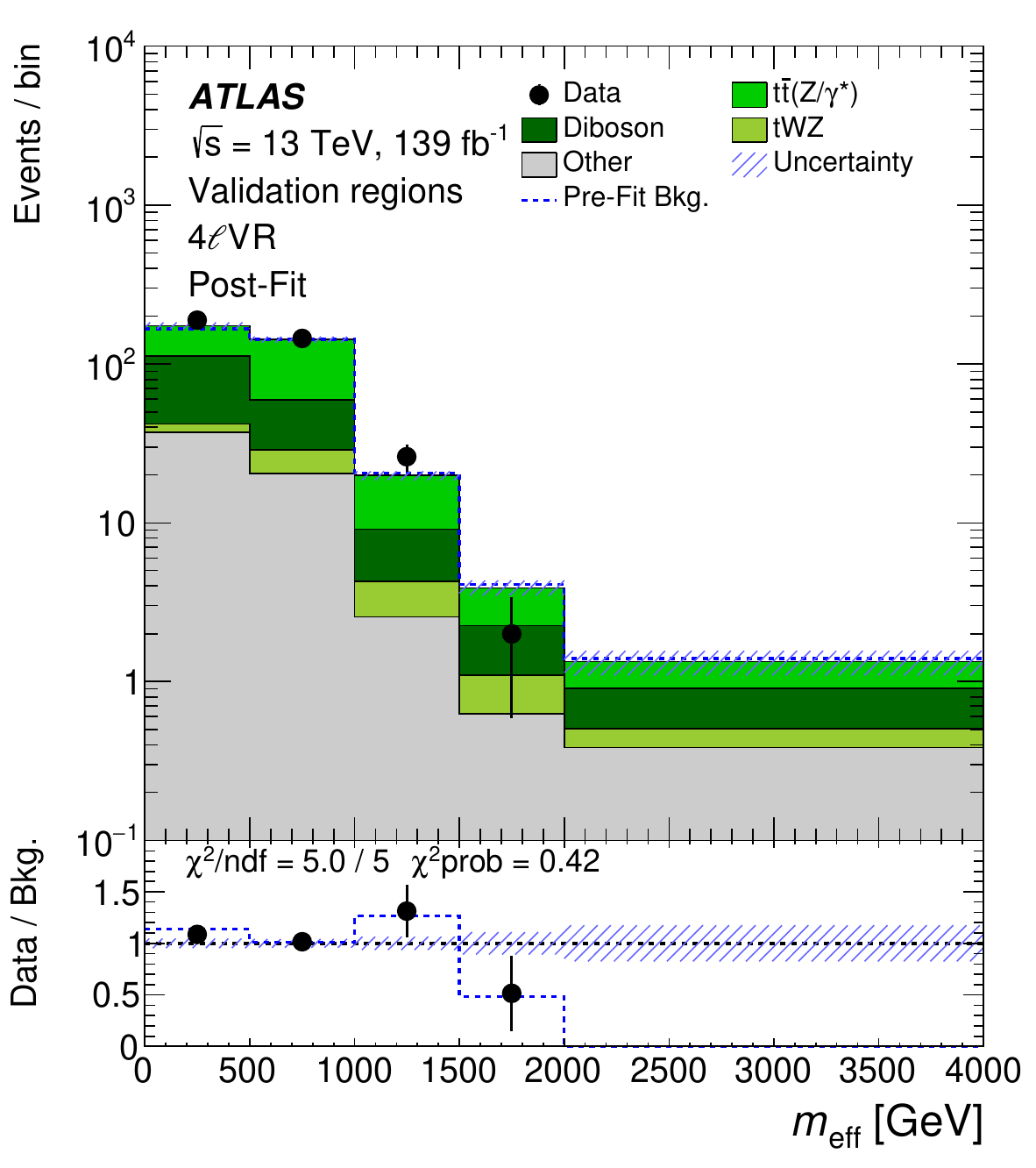}}\\
\caption{
Comparison between data and prediction for the $\Meff$ distribution used in the validation regions: (a) $\lll$VR, (b) $\llll$VR.
The background contributions after the likelihood fit to data (``Post-Fit'') under the background-only hypothesis are shown as filled histograms.
The total background prediction before the likelihood fit to data (``Pre-Fit Bkg.'') is shown as a dashed blue histogram in the upper panel.
The ratio of the data to the background (``Bkg.'') prediction is shown in the lower panel, separately for post-fit background (black points) and pre-fit background (dashed blue line).
The size of the combined statistical and systematic uncertainty in the background prediction is indicated by the blue hatched band.
The last bin in each figure contains the overflow.
}
\label{fig:meffspectrum_VRs}
\end{figure*}
\clearpage


\section{Systematic uncertainties}
\label{sec:systematics}

Systematics uncertainties from the reconstructed objects and from the theoretical modelling of various physics processes can impact the estimated signal and background rates and the shape of the fitted distributions in multiple event categories.
The impact of systematic uncertainties in the fit results were found to be very small compared to the statistical uncertainties due to limited number of data events in the signal regions.
 
The sources of uncertainties due to instrumental effects include data luminosity and pile-up reweighting of reconstructed events in the MC simulations. The uncertainty in the combined 2015--2018 integrated luminosity is $1.7\%$~\cite{ATLAS-CONF-2019-021}, obtained using the LUCID-2 detector~\cite{LUCID2} for the primary luminosity measurements.  There are also uncertainties associated with the leptons' and jets' energy or momentum scale and resolutions~\cite{PERF-2016-04,JETM-2018-05}.  Furthermore, the uncertainties in the scale factors to correct for the differences between data and MC simulation in trigger~\cite{TRIG-2018-05,TRIG-2018-01}, lepton reconstruction, identification and isolation~\cite{EGAM-2018-01,MUON-2018-03}, jet vertex tagging~\cite{PERF-2014-03,ATLAS-CONF-2017-065} and $b$-tagging~\cite{FTAG-2018-01,FTAG-2020-08} efficiencies are taken into account. Some of these uncertainties are split into several components. These uncertainties are treated as correlated across the signal and background in the fit, but uncorrelated among individual components.
 
The impact of QCD renormalisation ($\mu_{R}$) and factorisation ($\mu_{F}$) scale variations by a factor of 0.5 or 2 with respect to the nominal scales, initial-state radiation scale variations and the $\alpha_s$ and \verb|NNPDF3.0nlo| PDFs~\cite{Butterworth:2015oua} are evaluated on the signal acceptance in the signal regions.  Total theoretical uncertainties on \LQmixd\ signal acceptance varies from $1\%$ to $10\%$ for low and intermediate masses and from $7\%$ to $19\%$ for high mass ($m_{\LQmixd} \geq 1.7$~TeV) depending on the signal region, with the dominant contributions arising from the initial-state radiation variations and the $\mu_{R}$ and $\mu_{F}$ scale variations. Similar uncertainties are found for the vector \vLQ\ signal acceptances in the electron and muon channels.
 
Theoretical uncertainties due to the modelling of various irreducible background are included in the analysis. Modelling uncertainties evaluated by comparing the fit distributions from nominal MC sample to those from the alternative MC generator predictions are estimated for the \ttW, \ttZ, and \ttH processes, as listed in Table~\ref{tab:mcconfig}.
Additional uncertainties are evaluated from renormalisation and factorisation scale variations by a factor of 0.5 and 2, relative to the nominal scales, for the \ttW, \ttZ, \ttH, and diboson samples.
 
The diboson background contribution with additional light jets has a 2\% uncertainty assigned on its cross section, as calculated from the largest statistical uncertainty associated to the data-driven diboson correction in the \lllVVnb region, whereas the normalisation of the $\VV$+HF background is determined from the likelihood fit to the data (see Section~\ref{sec:bkg_irreducible}).  Furthermore, an uncertainty derived from the fitted parameters for the jet-multiplicity-dependent diboson correction is additionally applied to diboson processes (see Section~\ref{sec:bkg_irreducible}). The diboson scale variation uncertainties are uncorrelated between LF and HF, whereas the jet-multiplicity-dependent diboson correction uncertainty is correlated across the LF and HF components.
 
Various sources of uncertainty in the estimation of the non-prompt leptons, photon conversions and charge misidentified background are taken into account.
An extrapolation uncertainty of 20\% is assigned to capture possible differences in mismodelling between the $T$ and the $M_{ex}$ lepton definitions. This uncertainty is estimated by comparing the $T$-over-$M$ non-prompt rate between the nominal \textsc{Powheg+Pythia8} \ttbar sample and the alternative \textsc{Powheg+Herwig7} \ttbar sample. The input variables to the non-prompt lepton BDT were studied and non-closure uncertainties were derived for three of them displaying some data and prediction disagreement in the \llSS $TM_{ex}$ region. These variables are the relative muon calorimeter cluster energy ($E_{\mathrm{cluster}}/E_{\mathrm{expected}}$), the electron track \pt divided by the jet track \pt, and the secondary vertex longitudinal significance using tracks with \pt $>$ 500 MeV for both electrons and muons. The statistical uncertainty associated to the data-driven correction of the $b$-jet multiplicity distribution in the non-prompt lepton background is considered as a separate uncertainty. Additionally, a 50\% uncertainty is assigned to the \ttbar+$\geq 1$ $b/c$-jet contribution, although the result is insensitive to this choice and remains unchanged if a 100\% uncertainty is assigned instead.
 
Uncertainties of 50\% and 10\% are estimated for internal and material conversions, respectively, based on a comparison between data and simulation in a dedicated \llSS control sample enhanced in conversion candidate events. This sample is defined by requiring two same-sign $T$ leptons, at least one of which is an electron that fails the internal or material conversion veto requirement, respectively. This uncertainty is applied to all categories, except for \lllintc and \lllmatc, as an extrapolation uncertainty.
 
The electron charge misassignment measurement is validated by a closure test in simulation using same-charge electron pairs, with the observed difference between measured and predicted rates taken as the systematic uncertainty. Additional systematic uncertainties include the statistical uncertainty from the data and the variation in the rates when the $Z$-peak range definition is varied. The total systematic uncertainty in the charge misassignment background estimate for $M$ electrons is about 30\%, with the dominant contribution at low $\pt$ originating from the closure tests and at high $\pt$ from the statistical uncertainty.
 
Uncertainties in the modelling of other rare processes (e.g $t\bar t t\bar t$, $t (Z/\gamma^*)$, $t W (Z/\gamma^*)$, $t\bar t t$, $t\bar t W^+ W^-$, $VH$, triboson) are taken into account by assigning up to $50\%$ uncertainties on the production cross sections. These are conservative uncertainties to account for limited information on the correct modelling of these background processes. The choice of the size of these uncertainties, as well as the \ttbar+$\geq 1$ $c$-jet uncertainties, have negligible impact on the results.


\section{Results}
\label{sec:results}

A maximum-likelihood fit is performed for each signal hypothesis (\tete\ and \tmtm) on all bins in the 7 control regions and in the corresponding 2 signal regions to simultaneously determine the background and the leptoquark signal yields that are most consistent with the data.
For the \tete\ (\tmtm) result, the $\lll$SR-$e$ and $\llll$SR-$e$ ($\lll$SR-$\mu$ and $\llll$SR-$\mu$) signal regions are considered in the fit. All control regions are included in each fit. In the signal regions, as well as in the \llttW, \lllttZ, and \lllVV control regions, the \Meff distribution is fitted, thus ensuring that the main SM background processes are correctly estimated in the tails of high \Meff values, as well as maximising the sensitivity to the leptoquark signal. In the remaining control regions \lltte, \llttmu, \lllintc, and \lllmatc, the total event yield (i.e. a single bin) is used.
 
The likelihood function ${\cal L}(\mu,\vec{\lambda}, \vec{\theta})$ is constructed as a product of Poisson probability terms over all bins considered in the search.  It depends on the signal-strength parameter, $\mu$, defined as a multiplicative factor applied to the predicted yield for the pair-produced leptoquark signal, $\vec{\lambda}$, the normalisation factors for several backgrounds (see Section~\ref{sec:background}), and $\vec{\theta}$, a set of nuisance parameters (NP) encoding systematic uncertainties in the signal and background expectations~\cite{Cranmer:2012sba}. The predicted signal yield depends on the assumed mass $m_{\LQ}$ and on the leptoquark being a scalar $\LQ$ or a vector $\vLQ$. Systematic uncertainties can impact the estimated signal and background rates, the migration of events between categories, and the shape of the fitted distributions; they are summarised in Section~\ref{sec:systematics}.
Both $\mu$ and $\vec{\lambda}$ are treated as free parameters in the likelihood fit. The NPs $\vec{\theta}$ allow variations of the expectations for signal and background according to the systematic uncertainties, subject to Gaussian constraints in the likelihood fit.
Their fitted values represent the deviations from the nominal expectations that globally provide the best fit to the data.
Statistical uncertainties in each bin due to the limited size of the simulated samples are taken into account by dedicated parameters using the Beeston--Barlow ``lite'' technique~\cite{Barlow:1993dm}.
 
The test statistic $q_\mu$ is defined as the profile likelihood ratio:
$q_\mu = -2\ln({\cal L}(\mu,{\hat{\vec{\lambda}}}_\mu, {\hat{\vec{\theta}}}_\mu)/{\cal L}(\hat{\mu},\hat{\vec{\lambda}}_{\hat{\mu}}, \hat{\vec{\theta}}_{\hat{\mu}}))$,
where $\hat{\mu}$, $\hat{\vec{\lambda}}_{\hat{\mu}}$, and $\hat{\vec{\theta}}_{\hat{\mu}}$ are the values of the parameters that maximise the likelihood function,
and ${\hat{\vec{\lambda}}}_\mu$ and ${\hat{\vec{\theta}}}_\mu$ are the values of the parameters that maximise the likelihood function for a given value of $\mu$.
The test statistic $q_\mu$ is evaluated with the {\textsc RooFit} package~\cite{Verkerke:2003ir}.
A related statistic is used to determine the probability that the observed data are compatible with the background-only hypothesis (i.e.~the discovery test) by setting $\mu=0$ in the profile likelihood ratio ($q_0$).
The $p$-value (referred to as $p_0$) representing the probability of the data being compatible with the background-only hypothesis is estimated by integrating the distribution of $q_0$ from background-only pseudo-experiments
above the observed value of $q_0$. Some model dependence exists in the estimation of the $p_0$, as a given signal scenario needs to be assumed in the calculation of the denominator of $q_0$, even if the overall signal normalisation is allowed to float and is fitted to data. The observed $p_0$ is checked for each explored signal scenario.
Upper limits on the signal production cross section for each of the signal scenarios considered are derived by using $q_\mu$ in the CL$_{\textrm{s}}$ method~\cite{Junk:1999kv,Read:2002hq}.
For a given signal scenario, values of the production cross section (parameterised by $\mu$) yielding CL$_{\textrm{s}} < 0.05$, where CL$_{\textrm{s}}$ is computed using pseudo-experiments,
are excluded at $\geq 95\%$ confidence level (CL).
 
\begin{figure*}[tbp]
\centering
\subfloat[]{\includegraphics[width=0.75\textwidth]{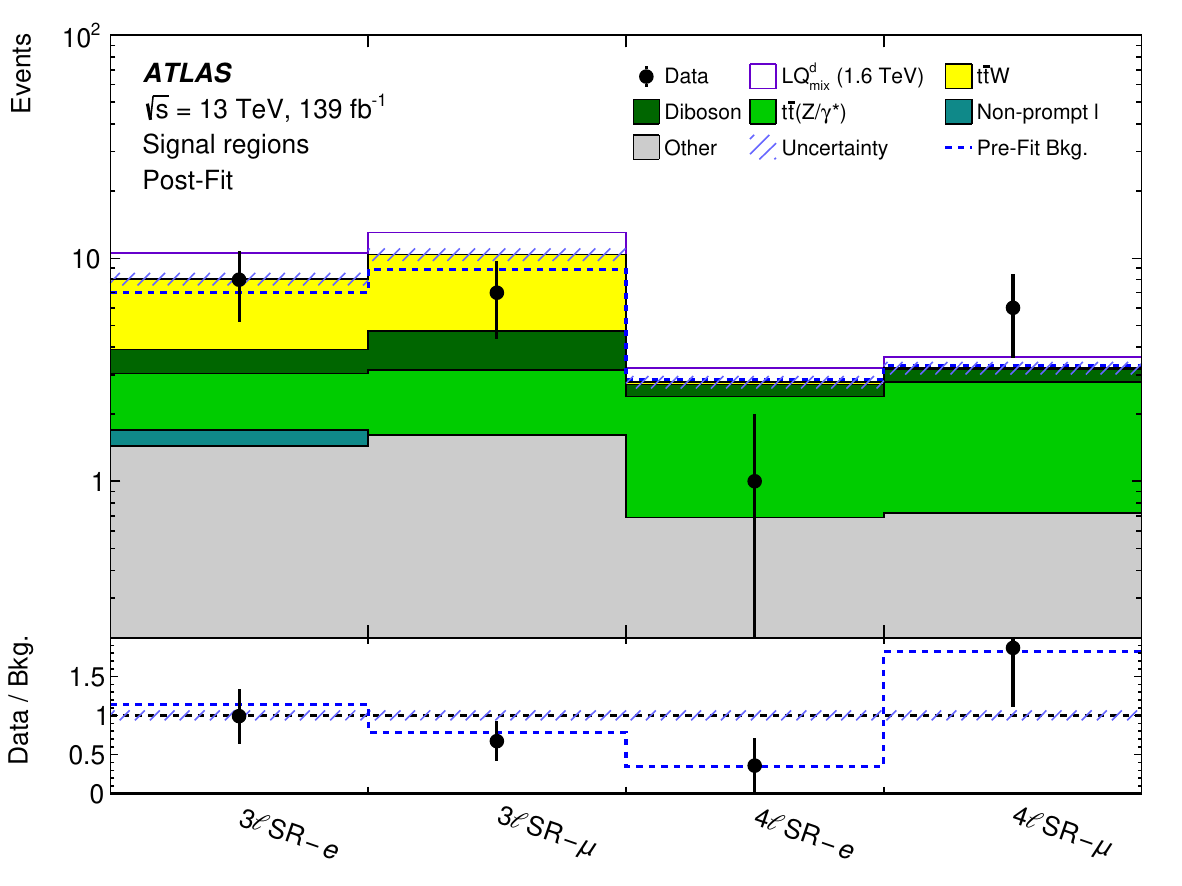}\label{fig:summarySR}} \\ 
\subfloat[]{\includegraphics[width=0.75\textwidth]{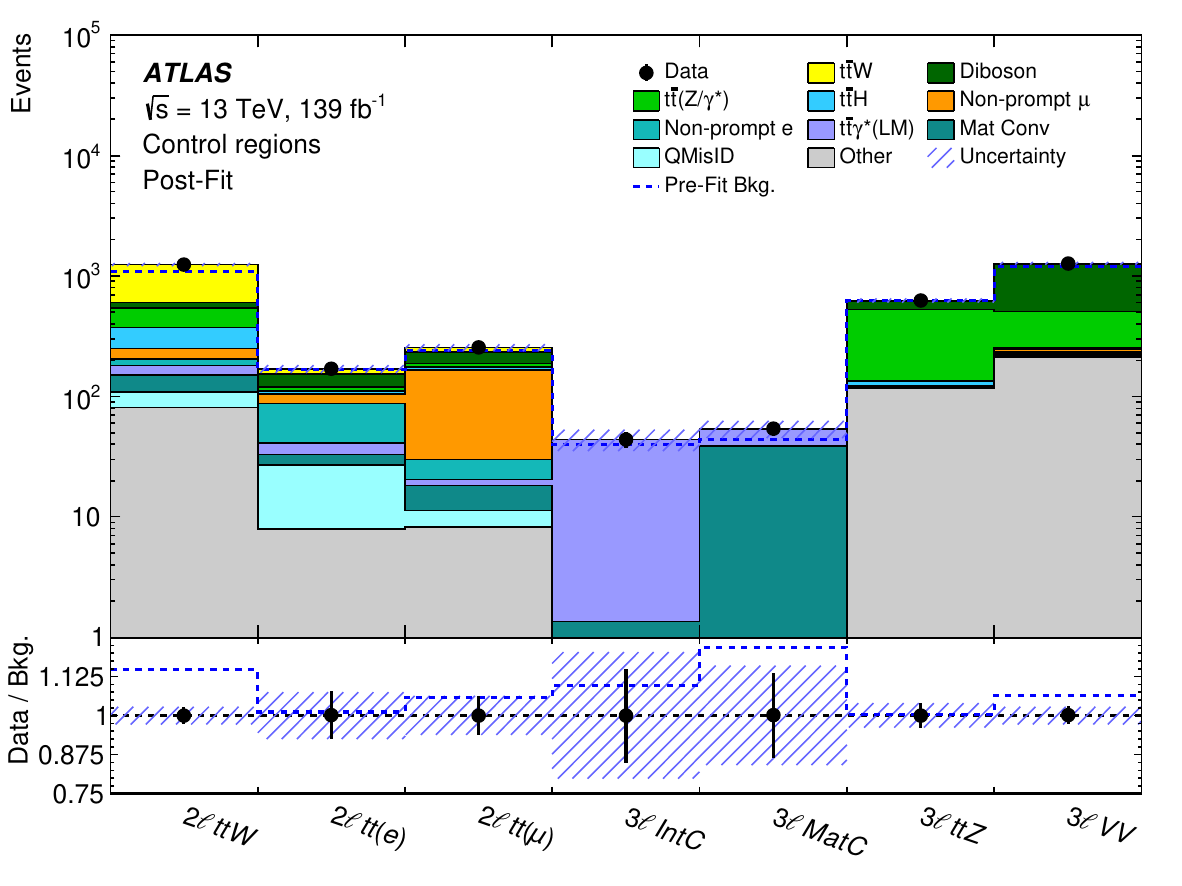}\label{fig:summaryCR}}
\caption{Comparison between data and the background prediction for the event yields in (a) the four signal region categories and (b) the seven control region categories.
The background contributions after the likelihood fit to data (``Post-Fit'') under the background-only hypothesis are shown as filled histograms. In (a) the expected \LQmixd\ ($m_{\LQmixd} = 1.6~\tev$) signal event yields before the fit are added to the background in the upper panel. The signal shown in \threelepel\ and \fourlepel\ (\threelepmu\ and \fourlepmu) SRs corresponds to \tete\ (\tmutmu).
The total background prediction before the likelihood fit to data (``Pre-Fit Bkg.'') is shown as a dashed blue histogram in the upper panel.
The ratio of the data to the background (``Bkg.'') prediction is shown in the lower panel, separately for post-fit background (black points) and pre-fit background (dashed blue line).
The size of the combined statistical and systematic uncertainty in the background prediction is indicated by the blue hatched band.
}
\end{figure*}
 
A comparison of the distributions of observed and expected yields in the four SRs and the seven CRs after the combined likelihood fit under the background-only hypothesis
is shown in Figures~\ref{fig:summarySR} and~\ref{fig:summaryCR}, respectively. The corresponding post-fit yields for the $\lll$ and $\llll$ SRs can be found in Table~\ref{tab:SRspostfityields}. The systematic uncertainty with the largest impact on the signal strength originates from the lepton identification. The search is dominated by statistical uncertainties.
In general, good agreement between the data and predicted background yields is found across all event categories.
 
\begin{table}[tbp]
\begin{center}
\caption{Summary of observed and predicted yields in the four signal region categories. The background prediction is shown after the combined likelihood fit to data under the background-only hypothesis across all control region and signal region categories. The expected signal yields that are obtained by using their theoretical cross sections are also shown with their pre-fit uncertainties, assuming $\BR$=1 and $\mu$=1. The uncertainties correspond to the combined statistical and systematic uncertainties in the predicted yields. The ``Other'' contribution is dominated by \tttt and \ttWW in the \lll SRs, whereas it is dominated by $tWZ$ and \ttWW in the \llll SRs. Dashes refer to components that are negligible or not applicable. \label{tab:SRspostfityields}}
\begin{tabular}{|l|c|c|c|c|}
\hline
\hline
& {3$\ell$SR-$e$} & {3$\ell$SR-$\mu$} & {4$\ell$SR-$e$} & {4$\ell$SR-$\mu$}\\
\hline
\hline
Data   & 8 & 7  & 1 & 6 \\
\hline
Total background & 8.1 $\pm$ 0.6 & 10.2 $\pm$ 0.7 & 2.8 $\pm$ 0.2 & 3.3 $\pm$ 0.2 \\
\hline
$t\bar{t}W$   & 4.2 $\pm$ 0.6 & 5.6 $\pm$ 0.8 & --- & --- \\
Diboson   & 0.9 $\pm$ 0.1 & 1.5 $\pm$ 0.2 & 0.32 $\pm$ 0.05 & 0.40 $\pm$ 0.04 \\
$t\bar{t}Z/\gamma^{*}$   & 1.33 $\pm$ 0.14 & 1.55 $\pm$ 0.15 & 1.69 $\pm$ 0.18 & 2.09 $\pm$ 0.21 \\
$tWZ$ & --- & ---  & 0.23 $\pm$ 0.12 & 0.22 $\pm$ 0.12\\
Non-prompt $\ell$   & 0.25 $\pm$ 0.16 & --- & --- & --- \\ 
Other   & 1.44 $\pm$ 0.22 & 1.61 $\pm$ 0.31 & 0.53 $\pm$ 0.10 & 0.54 $\pm$ 0.12 \\
\hline
\hline
$\LQmixd$ 1.6 TeV   & 2.5 $\pm$ 0.2 & 2.7 $\pm$ 0.2 & 0.42 $\pm$ 0.11 & 0.40 $\pm$ 0.05 \\
$\vLQm$ 1.6 TeV   & 4.5 $\pm$ 0.2 & 4.6 $\pm$ 0.3 & 0.7 $\pm$ 0.1 & 0.7 $\pm$ 0.1 \\
$\vLQy$ 1.6 TeV   & 27 $\pm$ 1 & 29 $\pm$ 2 & 4.4 $\pm$ 0.2 & 4.2 $\pm$ 0.3 \\
$\vLQy$ 2.0 TeV    & 2.0 $\pm$ 0.2 & 2.0 $\pm$ 0.2 & 0.31 $\pm$ 0.08 & 0.30 $\pm$ 0.03 \\
\hline
\hline
\end{tabular}
\end{center}
\end{table}

The comparison between data and the background prediction for the \Meff distributions used in the different SRs is shown in Figure~\ref{fig:meffspectrum}. The binning used for the \Meff distributions in the different SRs represents a compromise between preserving enough discrimination in the fit between the background and the signal for the different values of the $m_{\LQ}$ considered, and keeping the statistical uncertainty of the background prediction per bin well below 30\%. No significant excess is observed in any of the SRs. The observed $p_0$ is found to be consistent with the background-only hypothesis for all values of $m_{\LQ}$ considered. The observed and expected $p_0$ as a function of $m_{\LQ}$ are shown in Figure~\ref{fig:p0scan}, assuming the value of $\BR=1$. This illustrates the significant expected sensitivity of the search, which exceeds 5 standard deviations for $m_{\LQ}<1.5~\tev$ and 3 standard deviations for $m_{\LQ}<1.6~\tev$.

\begin{figure*}[t]
\centering
\subfloat[]{\includegraphics[width=0.45\textwidth]{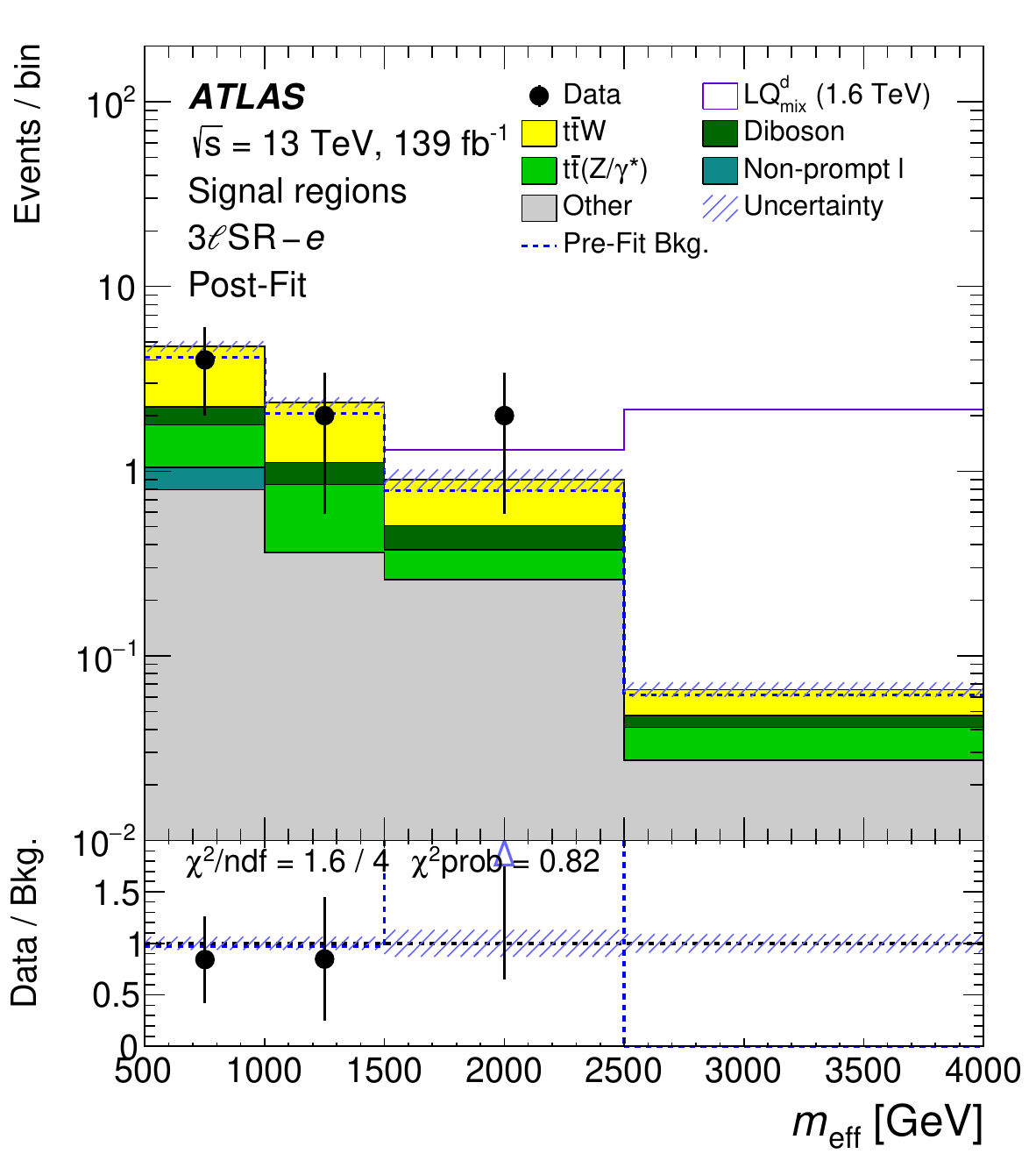}}
\subfloat[]{\includegraphics[width=0.45\textwidth]{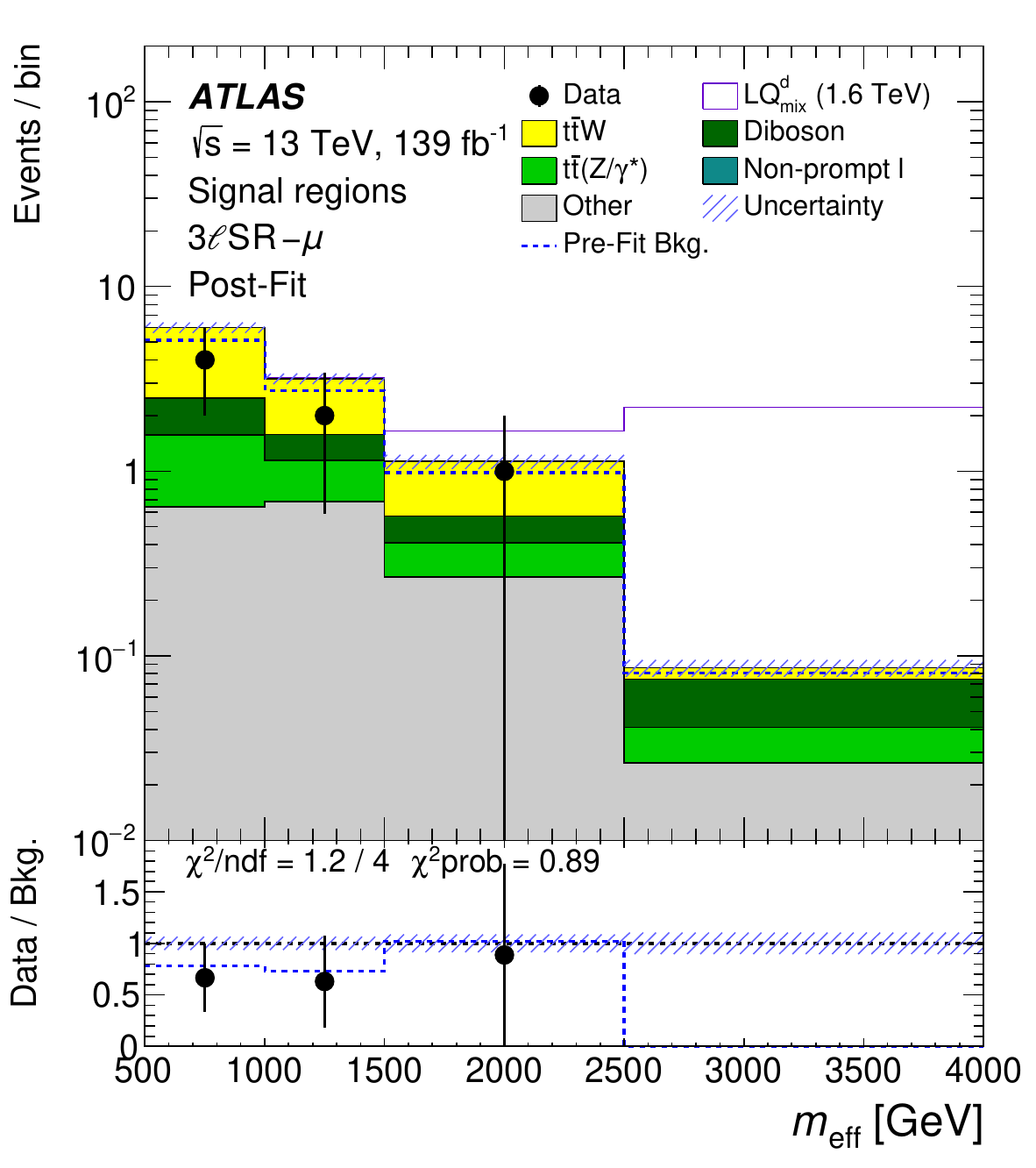}}\\
\subfloat[]{\includegraphics[width=0.45\textwidth]{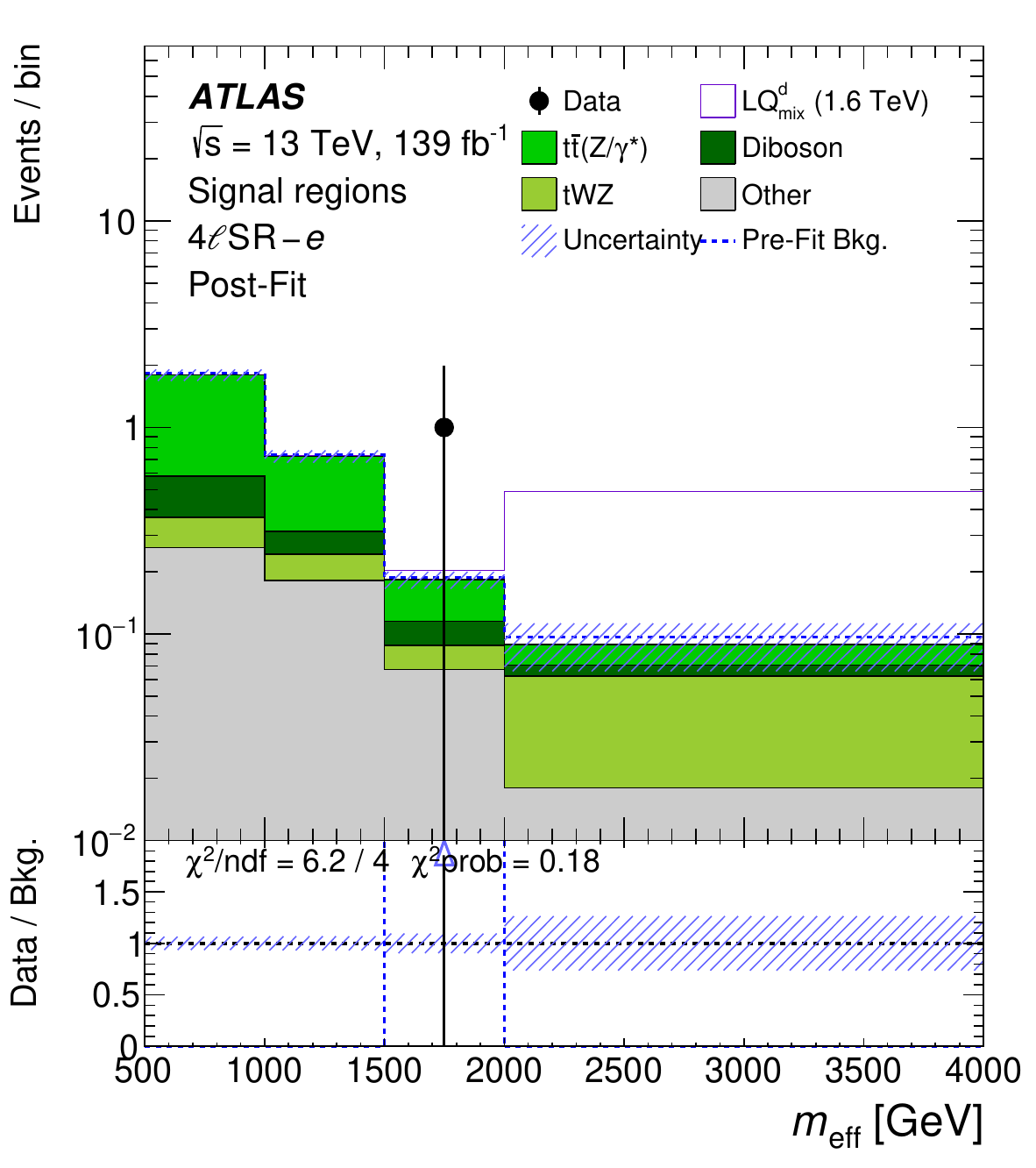}}
\subfloat[]{\includegraphics[width=0.45\textwidth]{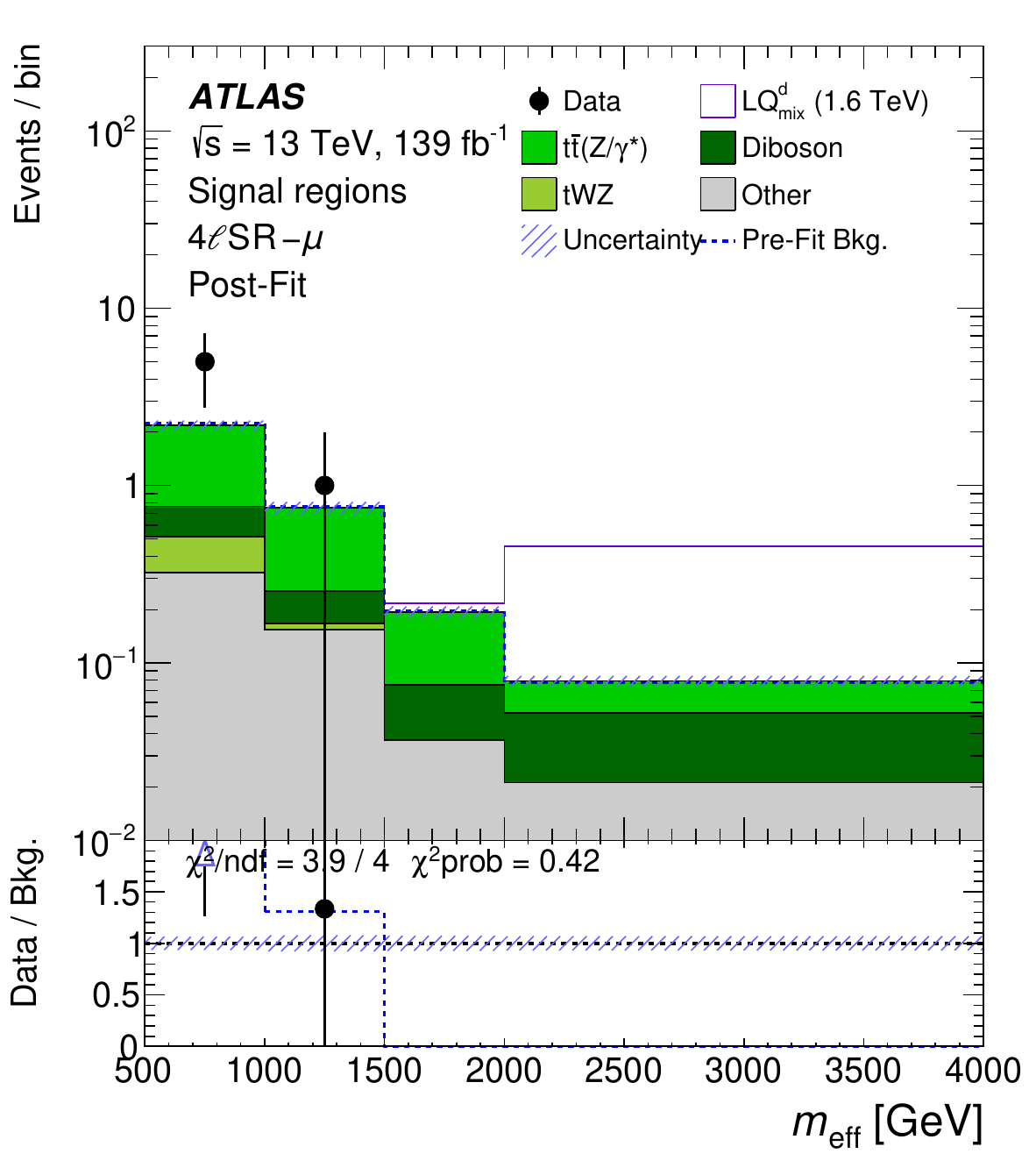}}
\caption{
Comparison between data and prediction for the $\Meff$ distribution used in different signal region categories:
(a) \threelepel, (b) \threelepmu, (c) \fourlepel, and (d) \fourlepmu.
The background contributions after the likelihood fit to data (``Post-Fit'') under the background-only hypothesis are shown as filled histograms.
The total background prediction before the likelihood fit to data (``Pre-Fit Bkg.'') is shown as a dashed blue histogram in the upper panel.
The ratio of the data to the background (``Bkg.'') prediction is shown in the lower panel, separately for post-fit background (black points) and pre-fit background (dashed blue line).
The expected \LQmixd\ ($m_{\LQmixd} = 1.6~\tev$) signal event yields before the fit are added to the background in the upper panel. The signal shown in \threelepel\ and \fourlepel\ (\threelepmu\ and \fourlepmu) SRs corresponds to \tete\ (\tmutmu).
The size of the combined statistical and systematic uncertainty in the background prediction is indicated by the blue hatched band.
The blue triangles indicate points that are outside the vertical range of the figure.
The last bin in each figure contains the overflow.
}
\label{fig:meffspectrum}
\end{figure*}
 
\begin{figure*}[t]
\centering
\includegraphics[width=0.7\textwidth]{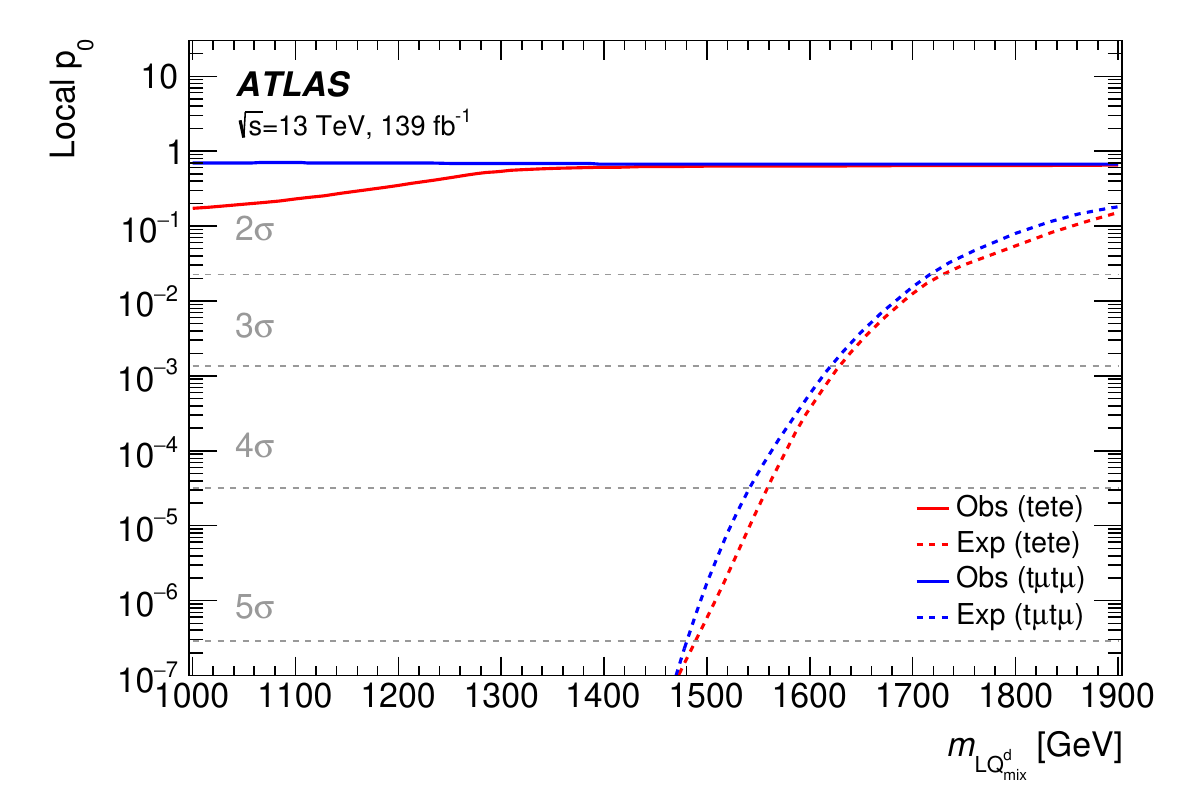}
\caption{The observed (solid) local $p_0$ as a function of $\LQ$ mass ($m_{\LQ}$) assuming $\BR=1$. The dashed curve shows the expected local $p_0$ under the hypothesis of a $\LQ$ signal at that mass. The horizontal dashed lines indicate the $p$-values corresponding to significances of 2 to 5 standard deviations.
}
\label{fig:p0scan}
\end{figure*}
 
In the absence of any significant excess above the SM background prediction, 95\% CL upper limits are set on the cross section for the $\LQ$ pair production as a function of the assumed $m_{\LQ}$ and $\BR$. Figures~\ref{fig:limitsLQtete} and~\ref{fig:limitsLQtmtm} show the 95\% CL upper limits on the $\LQ$ pair production cross section as a function of $m_{\LQ}$ for \tete\ and \tmtm, respectively, assuming $\BR=1$. The deviation in the observed \tete\ limit compared to the expected limit for masses below 1.6 TeV is caused by a slight upward fluctuation in data in the third bin of the \Meff distribution and by the leftward shift of the signal population in \Meff spectra with the decrease of $m_{\LQ}$. The uncertainty in the results is mainly due to limited statistics of selected data events. The size of the $\pm$1\,$\sigma$ uncertainty band around the expected limit shrinks for higher $\LQ$ mass hypotheses, for which the sensitivity is driven by \Meff bins with very low expected background yield (typically $\leq$ 0.1 events; see Figure~\ref{fig:meffspectrum}). This uncertainty band, which is computed using pseudo-experiments, has been validated by comparing two independent statistical frameworks.

Assuming $\BR=1$, the observed and expected 95\% CL lower limits on scalar $m_{\LQ}$ for \tete\ (\tmtm) are $1.58\ (1.59)~\tev$ and $1.59\ (1.59)~\tev$, respectively.
Exclusion limits on the scalar $m_{\LQ}$ are also obtained for various values of $\BR$ into a charged lepton and top quark (as shown in Figures~\ref{fig:limit2dte} and \ref{fig:limit2dtmu}).  As the signal search, with the requirement of at least three leptons, is only sensitive to $100\%$ \LQmixd $\rightarrow t\ell$ decay modes, the limits for the alternative $\BR$s, in steps of 0.1, are obtained by simply scaling theoretical cross sections by $\BR^2$ covering the full plane. Under the assumption of $50\%$ $\BR(\LQmixd \rightarrow te)$, the lower observed (expected) limit on the \LQ is 1.31 (1.35) TeV. \ In the case of $50\%$ $\BR(\LQmixd \rightarrow t\mu)$, the lower observed (expected) limit on the \LQ is 1.37 (1.36) TeV. \ These limits are competitive with the previous ATLAS limits established by the complementary dedicated search for \LQ in the mixed decay mode ($\LQ \rightarrow b\nu / t\ell$) with $50\%$ $\BR$~\cite{EXOT-2019-12}.

Additionally, 95\% CL upper limits are set on the $\vLQ$ pair production cross section as a function of $m_{\vLQ}$ in the minimal coupling scenario for \tete\ and \tmtm, as shown in Figures~\ref{fig:limitvLQteteMC} and~\ref{fig:limitvLQtmtmMC}, and in the Yang-Mills coupling scenario for \tete\ and \tmtm, as shown in Figures~\ref{fig:limitvLQteteYM} and~\ref{fig:limitvLQtmtmYM}.
Assuming $\BR=1$, the observed and expected 95\% CL lower limits on $m_{\vLQm}$ for minimum coupling scenario in \tete\ (\tmtm) are $1.67\ (1.67)~\tev$ and $1.67\ (1.67)~\tev$, respectively, whereas analogous lower limits on $m_{\vLQy}$ for Yang-Mills coupling scenario in \tete\ (\tmtm) are $1.95\ (1.95)~\tev$ and $1.95\ (1.95)~\tev$, respectively.

\begin{figure*}[t]
\centering
\subfloat[]{\includegraphics[width=0.7\textwidth]{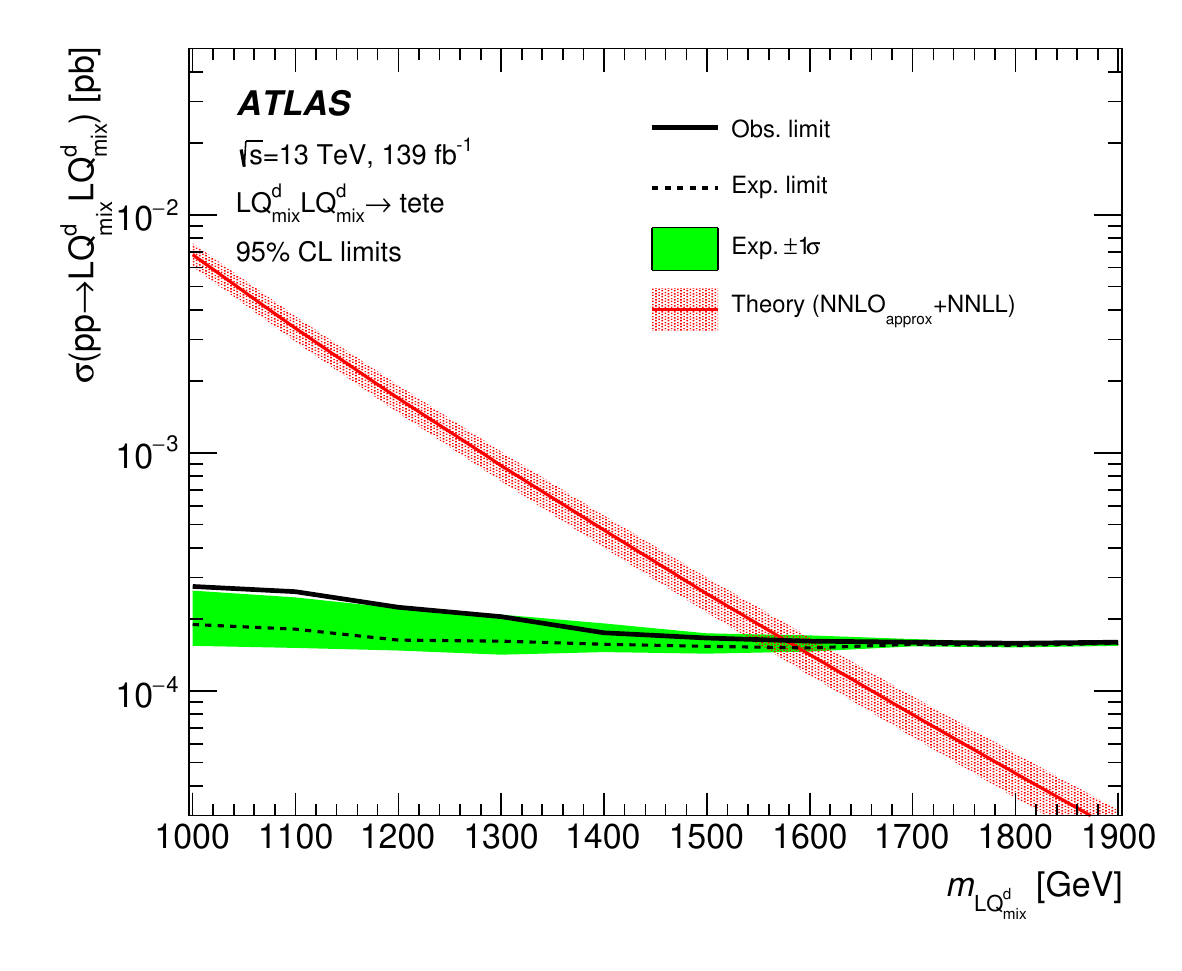}\label{fig:limitsLQtete}}\\
\subfloat[]{\includegraphics[width=0.7\textwidth]{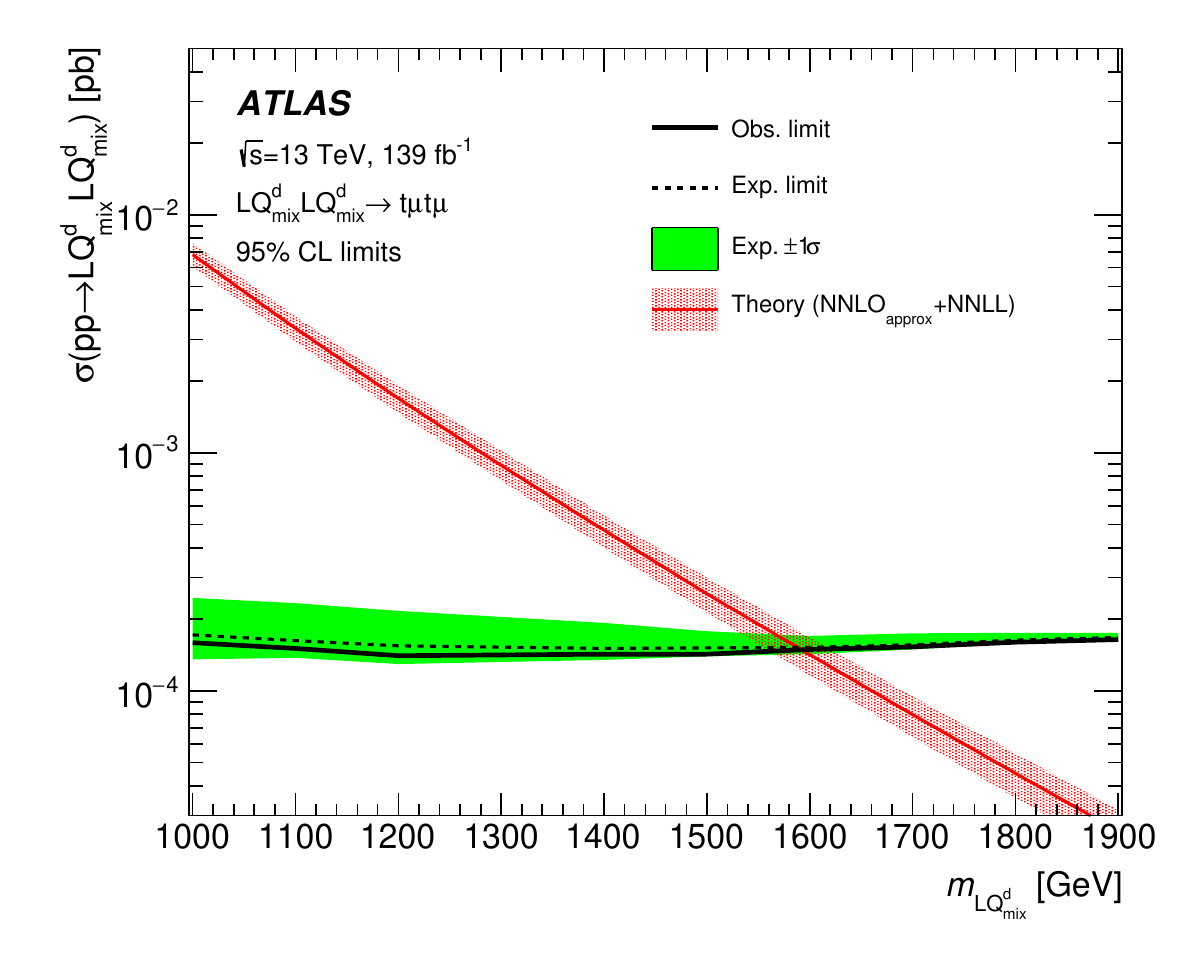}\label{fig:limitsLQtmtm}}
\caption{Observed (solid line) and expected (dashed line) 95\% CL upper limits on the $\LQ$ pair production cross section as a function of $m_{\LQ}$ for (a) \tete\ and (b) \tmtm.
The surrounding shaded band corresponds to the $\pm$1 standard deviation around the combined expected limit.
The red line and band show the theoretical prediction and its $\pm$1\,$\sigma$ uncertainty.
}
\end{figure*}
 
\begin{figure*}[t]
\centering
\subfloat[]{\includegraphics[width=0.7\textwidth]{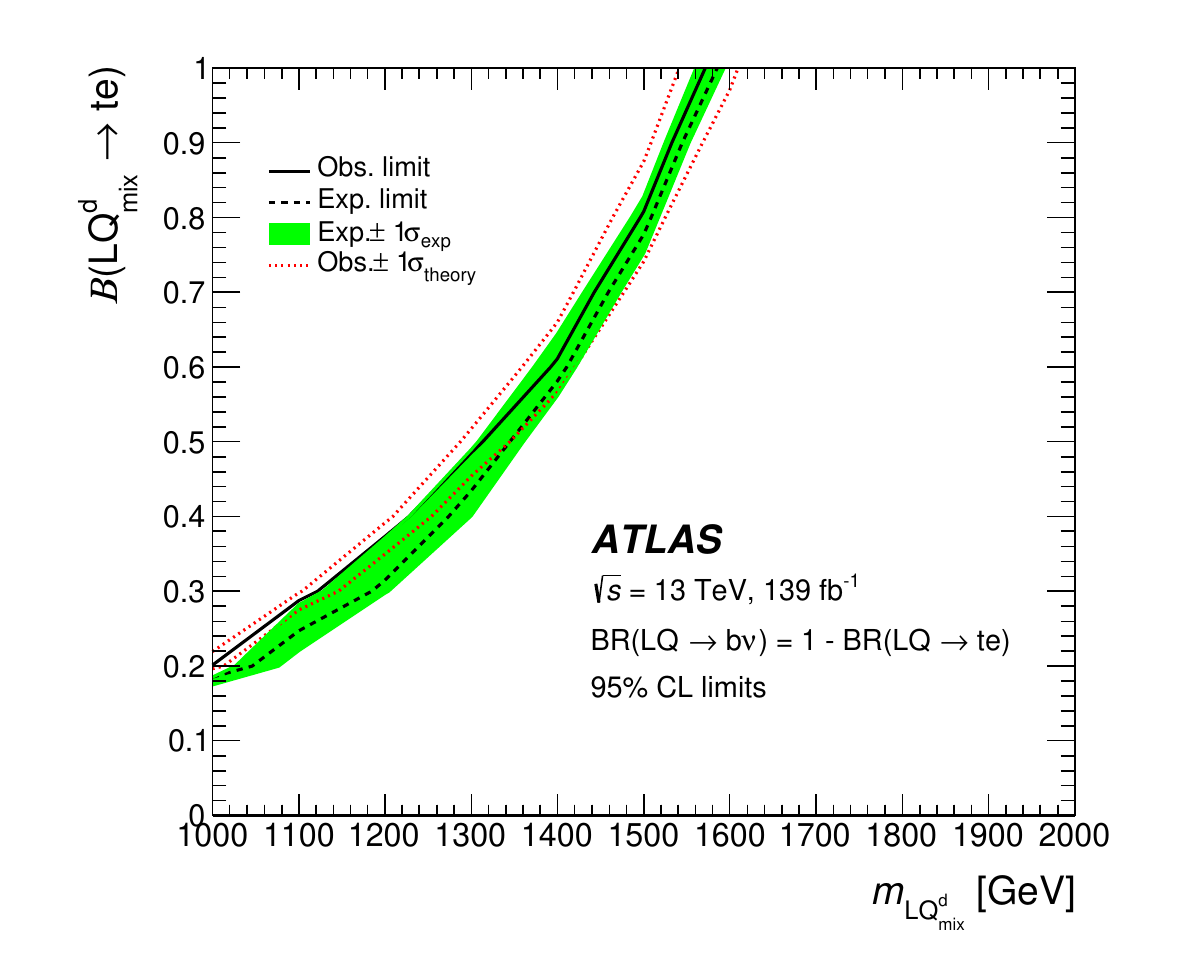}\label{fig:limit2dte}}\\
\subfloat[]{\includegraphics[width=0.7\textwidth]{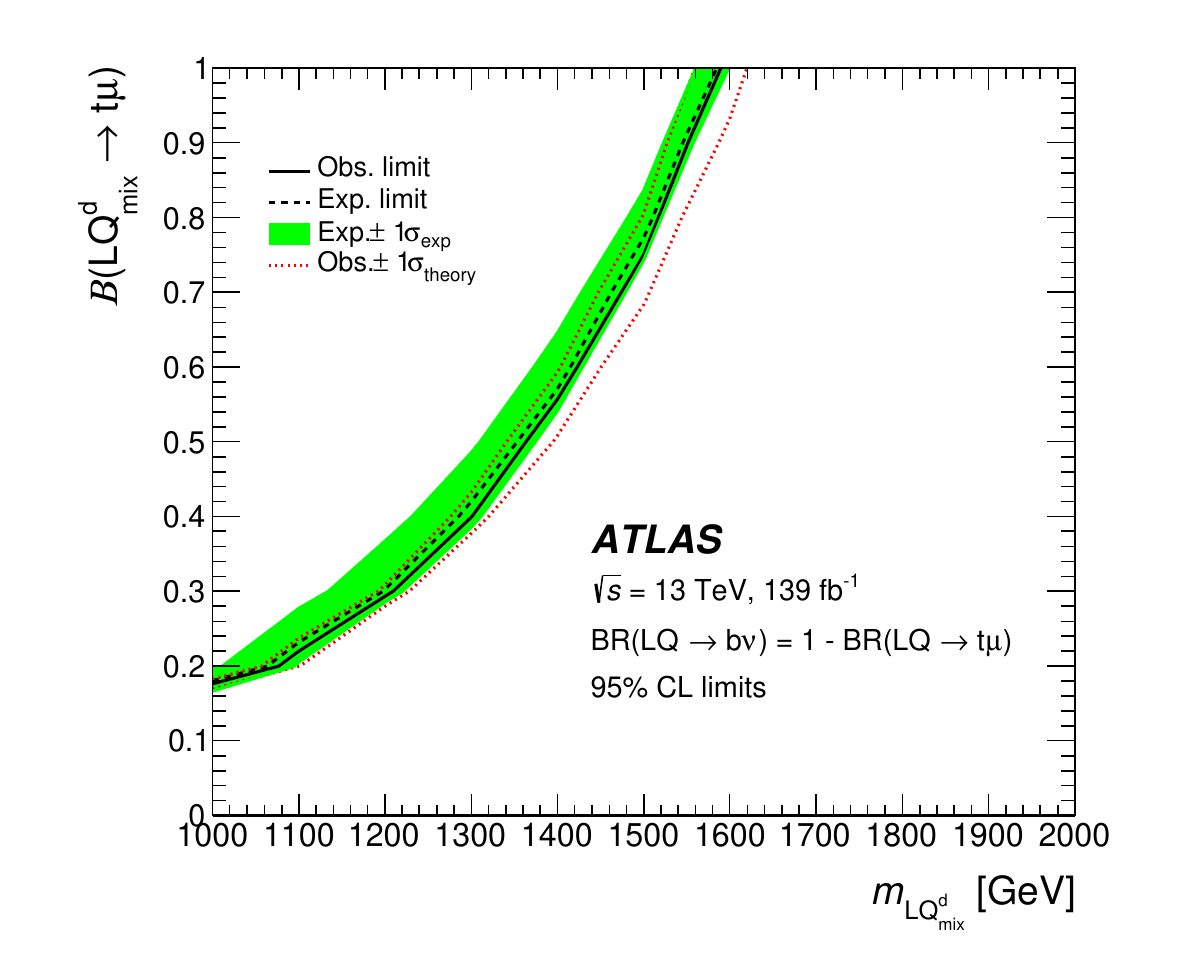}\label{fig:limit2dtmu}}
\caption{
Observed (solid line) and expected (dashed line) 95\% CL upper limits on $\BR$ as a function of $m_{\LQ}$, (a) limits resulting from the combination of \tete\ analysis channels, (b) limits resulting from the combination of \tmutmu\ analysis channels.  In each analysis, it is assumed that $\BR(\LQ \rightarrow b\nu) = 1 - \BR(\LQ \rightarrow t\ell)$, and there is no acceptance for $\LQ \rightarrow b\nu$ decays.  The surrounding shaded band corresponds to the $\pm$1\,$\sigma$ uncertainty around the combined expected limit, as estimated using pseudo-experiments.
The dotted red line around the observed limit indicates how the observed limit changes when varying the theoretical prediction for the $\LQ$ pair production cross section by its $\pm$1\,$\sigma$ uncertainty.
}
\end{figure*}
 
\begin{figure*}[t]
\centering
\subfloat[]{\includegraphics[width=0.50\textwidth]{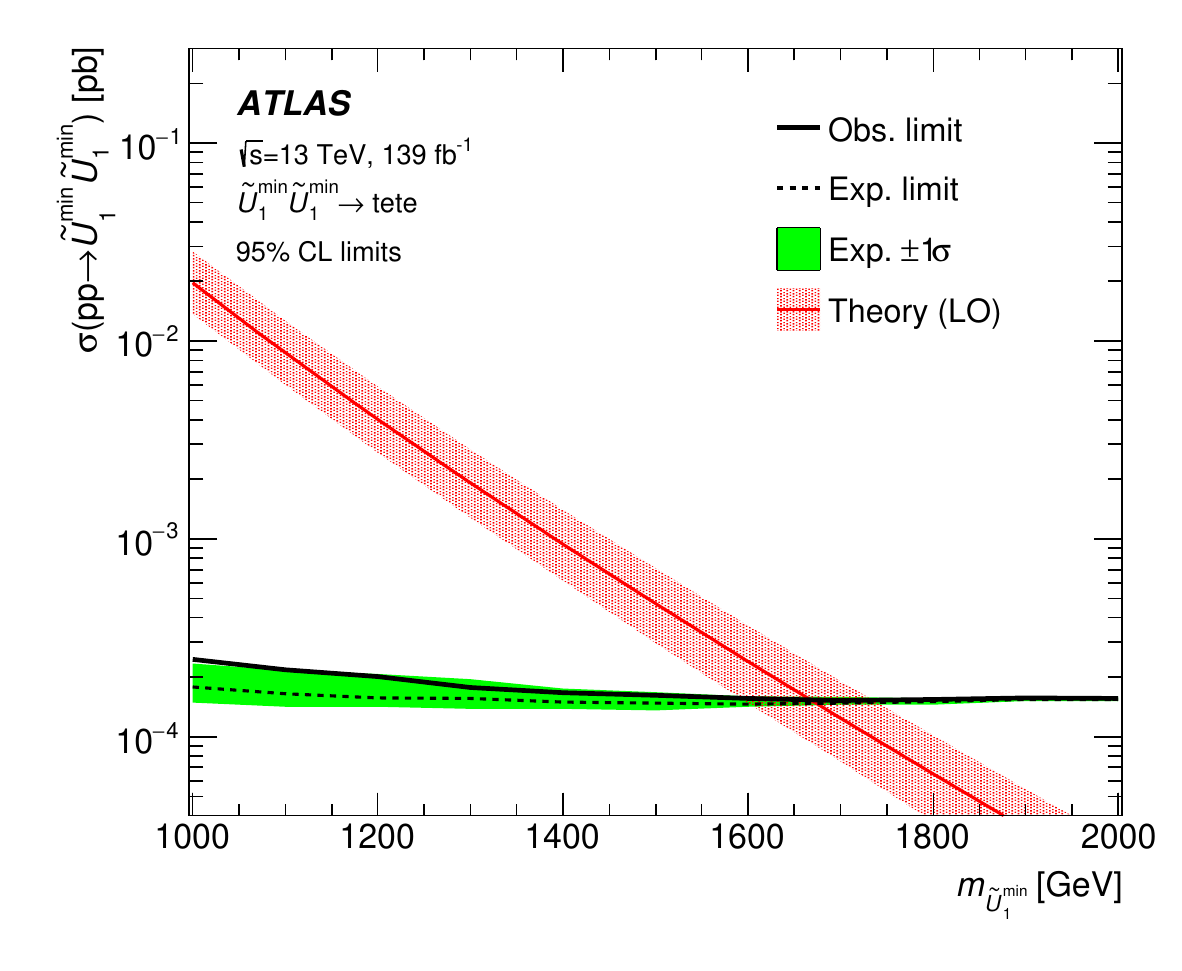}\label{fig:limitvLQteteMC}}
\subfloat[]{\includegraphics[width=0.50\textwidth]{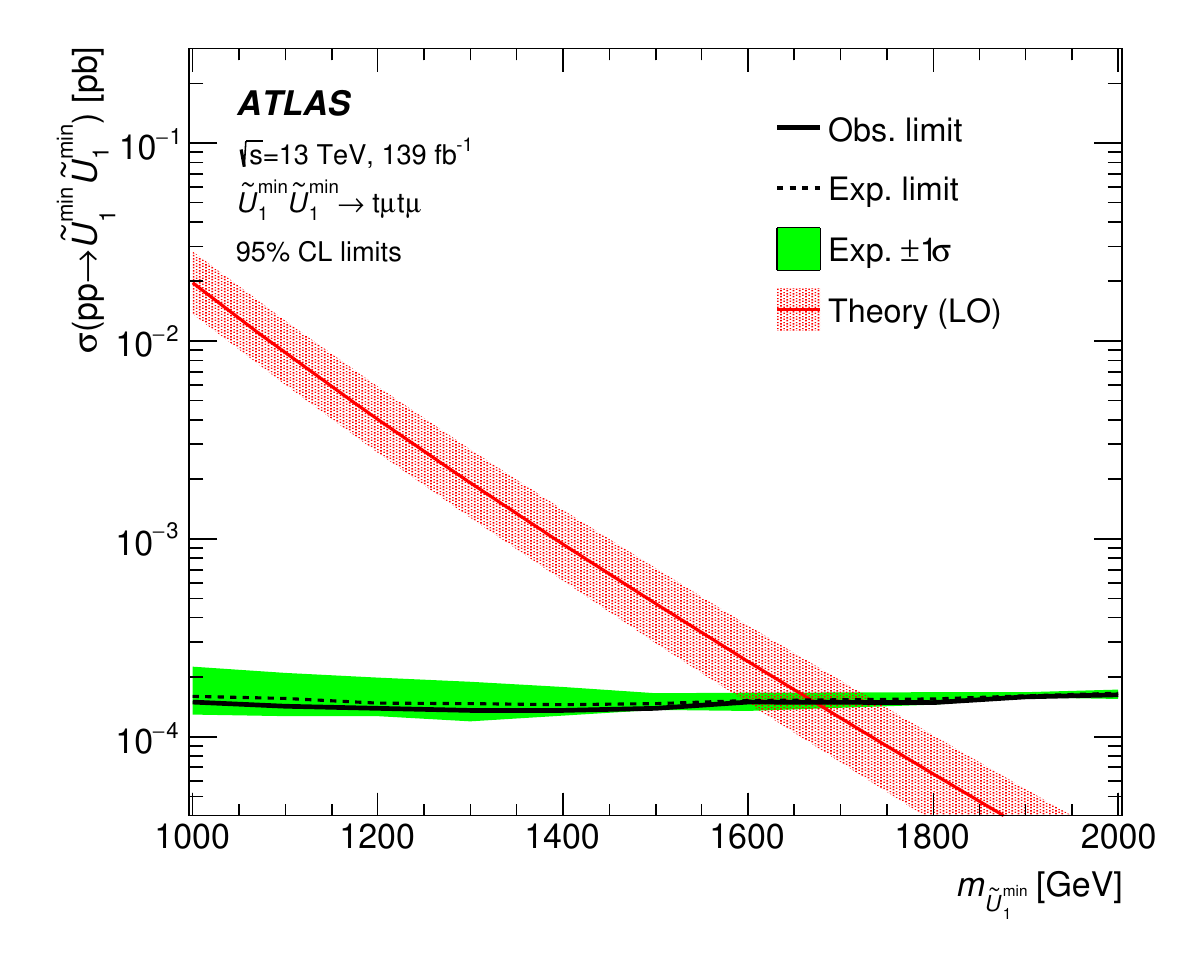}\label{fig:limitvLQtmtmMC}} \\
\subfloat[]{\includegraphics[width=0.50\textwidth]{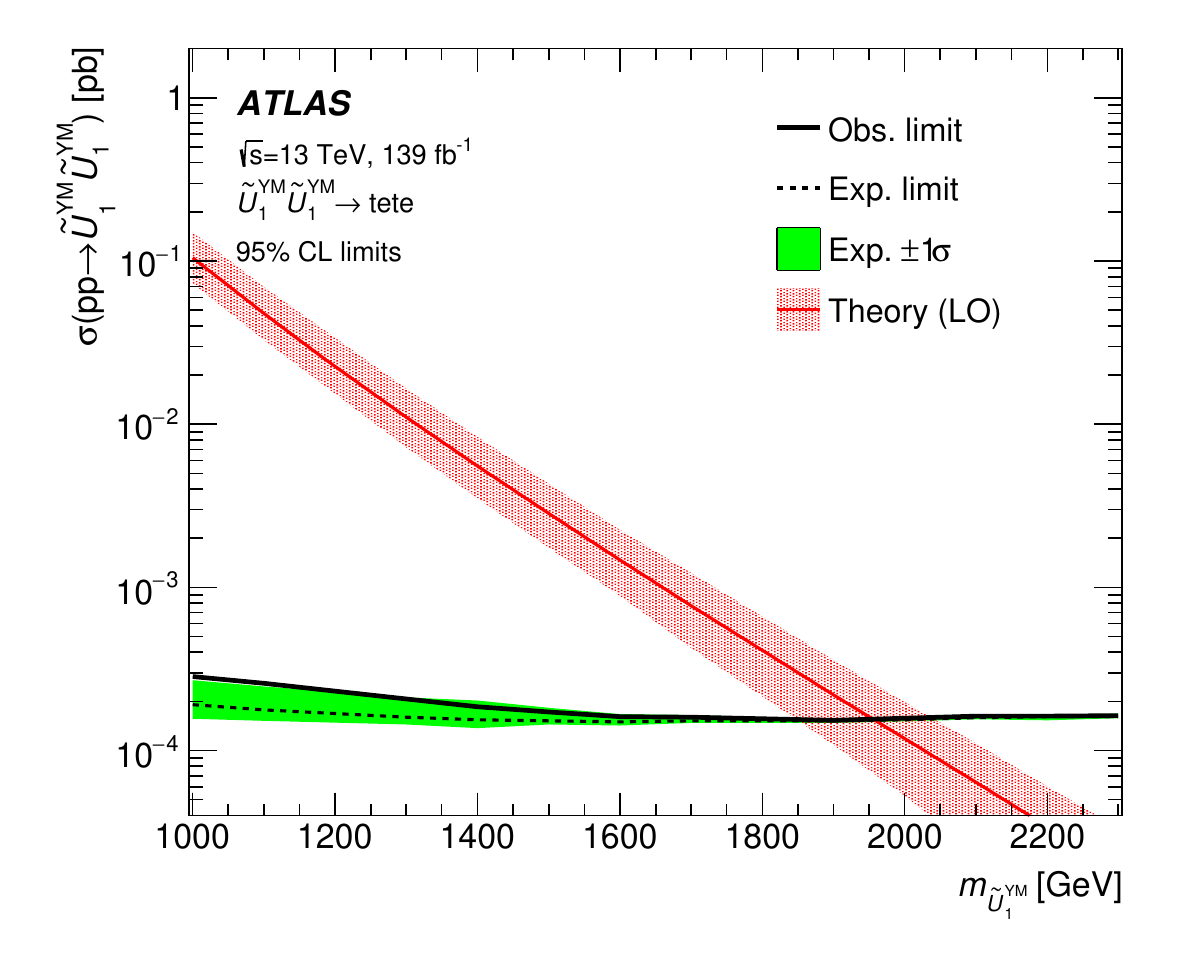}\label{fig:limitvLQteteYM}}
\subfloat[]{\includegraphics[width=0.50\textwidth]{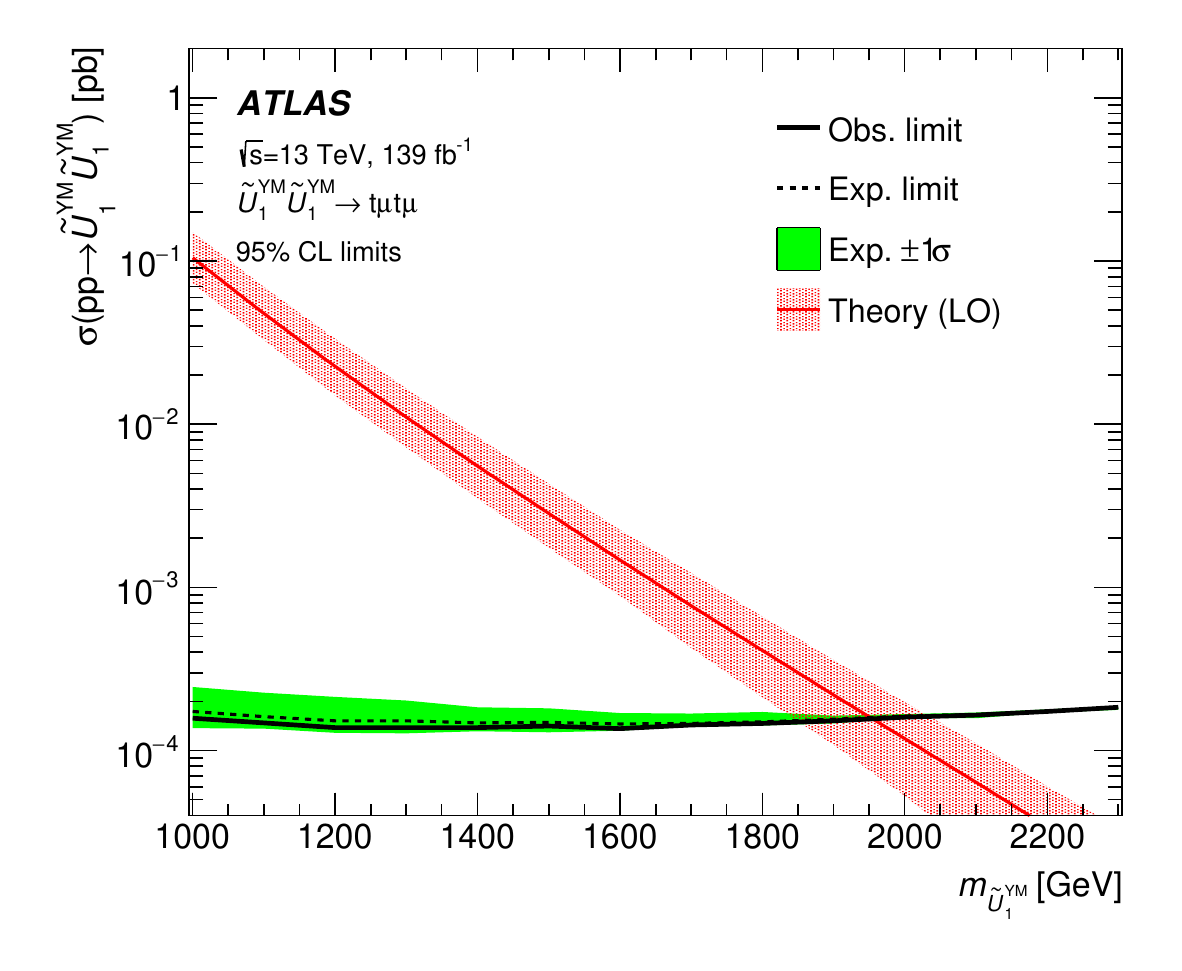}\label{fig:limitvLQtmtmYM}} \\
\caption{Observed (solid line) and expected (dashed line) 95\% CL upper limits on the $\vLQ$ pair production cross section as a function of $m_{\vLQ}$ for (a) minimal coupling scenario and \vLQm\ exclusive decay into $te$, (b) minimal coupling scenario and \vLQm\ exclusive decay into $t\mu$, (c) Yang-Mills scenario and $\vLQy$ exclusive decay into $te$, and (d) Yang-Mills scenario and $\vLQy$ exclusive decay into $t\mu$.
The surrounding shaded band corresponds to the $\pm$1 standard deviation around the combined expected limit.
The red line and band show the theoretical prediction and its $\pm$1\,$\sigma$ uncertainty.
}
\end{figure*}


\FloatBarrier
 
\section{Conclusion}

A search for pair production of cross-generation scalar and vector leptoquarks with a significant branching fraction into a top quark and an electron or muon is presented.
The search is based on the full Run~2 dataset recorded with the ATLAS detector at the Large Hadron Collider, which corresponds to 139~fb$^{-1}$ of $pp$ collisions at $\sqrt{s}=13~\tev$.
Events are selected if they have at least two light leptons (electrons or muons), and additional jets.
Four signal-enriched categories, defined by the multiplicity and flavour of the lepton candidates, are considered in the analysis in order to search for the leptoquark signal. Additional background-enriched categories are used in the fit to improve the modelling of several leading backgrounds.
The signal-enriched categories require at least three light leptons and employ the total effective mass distribution to discriminate between the signal and the background.
The search reaches an expected significance of 5 standard deviations for a scalar leptoquark decaying exclusively into $t\ell$ and with mass below about $1.5~\tev$. Its sensitivity is improved compared to previous searches with multiple light-leptons in the final state and is also complementary to searches performed with hadronically-decaying top quarks.
This results from the sophisticated event selection and categorisation employed, which ensures a high signal acceptance and low background yields.
No significant excess above the Standard Model expectation is observed in any of the considered event categories, and 95\% CL upper limits are set on the production cross section as a function of the leptoquark mass, for different assumptions of its nature. Scalar leptoquarks decaying exclusively into $te$ ($t\mu$) are excluded up to masses of $1.58\ (1.59)~\tev$. For the vector leptoquarks, the lower mass limit is about $1.67\ (1.67)~\tev$ in the minimal coupling scenario, and about $1.95\ (1.95)~\tev$ in the Yang-Mills coupling scenario for the decays into $te$ ($t\mu$), respectively.


\section*{Acknowledgements}


We thank CERN for the very successful operation of the LHC and its injectors, as well as the support staff at
CERN and at our institutions worldwide without whom ATLAS could not be operated efficiently.
 
The crucial computing support from all WLCG partners is acknowledged gratefully, in particular from CERN, the ATLAS Tier-1 facilities at TRIUMF/SFU (Canada), NDGF (Denmark, Norway, Sweden), CC-IN2P3 (France), KIT/GridKA (Germany), INFN-CNAF (Italy), NL-T1 (Netherlands), PIC (Spain), RAL (UK) and BNL (USA), the Tier-2 facilities worldwide and large non-WLCG resource providers. Major contributors of computing resources are listed in Ref.~\cite{ATL-SOFT-PUB-2023-001}.
 
We gratefully acknowledge the support of ANPCyT, Argentina; YerPhI, Armenia; ARC, Australia; BMWFW and FWF, Austria; ANAS, Azerbaijan; CNPq and FAPESP, Brazil; NSERC, NRC and CFI, Canada; CERN; ANID, Chile; CAS, MOST and NSFC, China; Minciencias, Colombia; MEYS CR, Czech Republic; DNRF and DNSRC, Denmark; IN2P3-CNRS and CEA-DRF/IRFU, France; SRNSFG, Georgia; BMBF, HGF and MPG, Germany; GSRI, Greece; RGC and Hong Kong SAR, China; ISF and Benoziyo Center, Israel; INFN, Italy; MEXT and JSPS, Japan; CNRST, Morocco; NWO, Netherlands; RCN, Norway; MEiN, Poland; FCT, Portugal; MNE/IFA, Romania; MESTD, Serbia; MSSR, Slovakia; ARRS and MIZ\v{S}, Slovenia; DSI/NRF, South Africa; MICINN, Spain; SRC and Wallenberg Foundation, Sweden; SERI, SNSF and Cantons of Bern and Geneva, Switzerland; MOST, Taipei; TENMAK, T\"urkiye; STFC, United Kingdom; DOE and NSF, United States of America.
 
Individual groups and members have received support from BCKDF, CANARIE, CRC and DRAC, Canada; PRIMUS 21/SCI/017 and UNCE SCI/013, Czech Republic; COST, ERC, ERDF, Horizon 2020 and Marie Sk{\l}odowska-Curie Actions, European Union; Investissements d'Avenir Labex, Investissements d'Avenir Idex and ANR, France; DFG and AvH Foundation, Germany; Herakleitos, Thales and Aristeia programmes co-financed by EU-ESF and the Greek NSRF, Greece; BSF-NSF and MINERVA, Israel; Norwegian Financial Mechanism 2014-2021, Norway; NCN and NAWA, Poland; La Caixa Banking Foundation, CERCA Programme Generalitat de Catalunya and PROMETEO and GenT Programmes Generalitat Valenciana, Spain; G\"{o}ran Gustafssons Stiftelse, Sweden; The Royal Society and Leverhulme Trust, United Kingdom.
 
In addition, individual members wish to acknowledge support from CERN: European Organization for Nuclear Research (CERN DOCT); Chile: Agencia Nacional de Investigación y Desarrollo (FONDECYT 1190886, FONDECYT 1210400); China: National Natural Science Foundation of China (NSFC - 12175119); EU: H2020 European Research Council (H2020-MSCA-IF-2020: HPOFHIC - 10103); European Union: European Research Council (ERC - 948254), Horizon 2020 Framework Programme (MUCCA - CHIST-ERA-19-XAI-00), Marie Sklodowska-Curie Actions (EU H2020 MSC IF GRANT NO 101033496); France: Agence Nationale de la Recherche (ANR-20-CE31-0013, ANR-21-CE31-0013, ANR-21-CE31-0022), Investissements d'Avenir Idex (ANR-11-LABX-0012); Germany: Deutsche Forschungsgemeinschaft (DFG - CR 312/5-1); Italy: Istituto Nazionale di Fisica Nucleare (FELLINI G.A. n. 754496); Japan: Japan Society for the Promotion of Science (JSPS KAKENHI 22H01227, JSPS KAKENHI JP21H05085, JSPS KAKENHI JP22H04944); Netherlands: Netherlands Organisation for Scientific Research (NWO Veni 2020 - VI.Veni.202.179); Norway: Research Council of Norway (RCN-314472); Poland: Polish National Agency for Academic Exchange (PPN/PPO/2020/1/00002/U/00001), Polish National Science Centre (NCN UMO-2019/34/E/ST2/00393, UMO-2020/37/B/ST2/01043, UMO-2021/40/C/ST2/00187); Slovenia: Slovenian Research Agency (ARIS grant J1-3010); Spain: BBVA Foundation (LEO22-1-603), Generalitat Valenciana (Artemisa, FEDER, IDIFEDER/2018/048), La Caixa Banking Foundation (La Caixa Foundation, LCF/BQ/PI20/11760025), Ministry of Science and Innovation (RYC2019-028510-I), PROMETEO and GenT Programmes Generalitat Valenciana (CIDEGENT/2019/023, CIDEGENT/2019/027, CIDEGENT/2019/029, GVA-SEJI/2020/037); Sweden: Swedish Research Council (SRC - 2017-05160, VR 2017-05092), Knut and Alice Wallenberg Foundation (KAW 2017.0100, KAW 2018.0157, KAW 2019.0447); Switzerland: Swiss National Science Foundation (SNSF - PCEFP2\_194658); UAE: Arab Fund for Economic and Social Development (AFESD-2021-994); United Kingdom: Leverhulme Trust (Leverhulme Trust RPG-2020-004).


\clearpage
\printbibliography
 
\clearpage
 
\begin{flushleft}
\hypersetup{urlcolor=black}
{\Large The ATLAS Collaboration}

\bigskip

\AtlasOrcid[0000-0002-6665-4934]{G.~Aad}$^\textrm{\scriptsize 101}$,
\AtlasOrcid[0000-0002-5888-2734]{B.~Abbott}$^\textrm{\scriptsize 119}$,
\AtlasOrcid[0000-0002-7248-3203]{D.C.~Abbott}$^\textrm{\scriptsize 102}$,
\AtlasOrcid[0000-0002-1002-1652]{K.~Abeling}$^\textrm{\scriptsize 55}$,
\AtlasOrcid[0000-0002-8496-9294]{S.H.~Abidi}$^\textrm{\scriptsize 29}$,
\AtlasOrcid[0000-0002-9987-2292]{A.~Aboulhorma}$^\textrm{\scriptsize 35e}$,
\AtlasOrcid[0000-0001-5329-6640]{H.~Abramowicz}$^\textrm{\scriptsize 150}$,
\AtlasOrcid[0000-0002-1599-2896]{H.~Abreu}$^\textrm{\scriptsize 149}$,
\AtlasOrcid[0000-0003-0403-3697]{Y.~Abulaiti}$^\textrm{\scriptsize 116}$,
\AtlasOrcid[0000-0003-0762-7204]{A.C.~Abusleme~Hoffman}$^\textrm{\scriptsize 136a}$
% ...
\bigskip
\\

% ...
$^{29}$Physics Department, Brookhaven National Laboratory, Upton NY; United States of America.\\
% ...
$^{35}$$^{(a)}$Facult\'e des Sciences Ain Chock, R\'eseau Universitaire de Physique des Hautes Energies - Universit\'e Hassan II, Casablanca;$^{(b)}$Facult\'{e} des Sciences, Universit\'{e} Ibn-Tofail, K\'{e}nitra;$^{(c)}$Facult\'e des Sciences Semlalia, Universit\'e Cadi Ayyad, LPHEA-Marrakech;$^{(d)}$LPMR, Facult\'e des Sciences, Universit\'e Mohamed Premier, Oujda;$^{(e)}$Facult\'e des sciences, Universit\'e Mohammed V, Rabat;$^{(f)}$Institute of Applied Physics, Mohammed VI Polytechnic University, Ben Guerir; Morocco.\\
% ...
$^{55}$II. Physikalisches Institut, Georg-August-Universit\"{a}t G\"ottingen, G\"ottingen; Germany.\\
% ...
$^{101}$CPPM, Aix-Marseille Universit\'e, CNRS/IN2P3, Marseille; France.\\
$^{102}$Department of Physics, University of Massachusetts, Amherst MA; United States of America.\\
% ...
$^{116}$Department of Physics, New York University, New York NY; United States of America.\\
% ...
$^{119}$Homer L. Dodge Department of Physics and Astronomy, University of Oklahoma, Norman OK; United States of America.\\
% ...
$^{136}$$^{(a)}$Departamento de F\'isica, Pontificia Universidad Cat\'olica de Chile, Santiago;$^{(b)}$Millennium Institute for Subatomic physics at high energy frontier (SAPHIR), Santiago;$^{(c)}$Instituto de Investigaci\'on Multidisciplinario en Ciencia y Tecnolog\'ia, y Departamento de F\'isica, Universidad de La Serena;$^{(d)}$Universidad Andres Bello, Department of Physics, Santiago;$^{(e)}$Instituto de Alta Investigaci\'on, Universidad de Tarapac\'a, Arica;$^{(f)}$Departamento de F\'isica, Universidad T\'ecnica Federico Santa Mar\'ia, Valpara\'iso; Chile.\\
% ...
$^{149}$Department of Physics, Technion, Israel Institute of Technology, Haifa; Israel.\\
$^{150}$Raymond and Beverly Sackler School of Physics and Astronomy, Tel Aviv University, Tel Aviv; Israel.\\
% ...
$^{a}$ Also Affiliated with an institute covered by a cooperation agreement with CERN.\\
$^{b}$ Also at An-Najah National University, Nablus; Palestine.\\
$^{c}$ Also at Borough of Manhattan Community College, City University of New York, New York NY; United States of America.\\
$^{d}$ Also at Bruno Kessler Foundation, Trento; Italy.\\
$^{e}$ Also at Center for High Energy Physics, Peking University; China.\\
$^{f}$ Also at Center for Interdisciplinary Research and Innovation (CIRI-AUTH), Thessaloniki; Greece.\\
% ...

\end{flushleft}


\clearpage
 
\appendix

\end{document}